# The Growth of Passive Membranes: Evidence from Ti and Like Metals

**Authors:** Weiwei Lao[1]†, Qiaojie Luo[1]†, Ying Huang[2]†, Haixu Zhong[3], Chaoqian Lou[1], Xuliang Deng[2]*, Xiufang Wen[3], Xiaodong Li[1]*

**Affiliations:**

[1]School of Stomatology, Stomatology Hospital, Zhejiang University School of Medicine, Hangzhou 310000, P. R. China.

[2]National Engineering Laboratory for Digital and Material Technology of Stomatology, Peking University School and Hospital of Stomatology, Beijing 100081, PR China.

[3]School of Chemistry and Chemical Engineering, South China University of Technology, Guangzhou, 510640, China.

*Corresponding author. Email: cisarli@zju.edu.cn; kqdengxuliang@bjmu.edu.cn

†These authors contributed equally to this work.

**Abstract:** Current contradictory understanding of passivation comes from overly-complex passive models, defective characterization and misplaced theoretical approaches. From brand-new experimentation, we find that a Ti passive membrane has spatiotemporally-ordered macrostructure. At the start, a thermodynamically-stable chemisorbed Ti-O monolayer is immediately formed to inactivate the outmost Ti atoms and shield direct reaction of environmental oxygens on metallic matrix, and then an underneath TiOx@Ti ceramet-like non-equilibrium gradient oxide layer rapidly forms. The two layers work synergistically to keep the macro-ordered passive membrane growing slowly via non-linear mechanism of incremental oxidation damping, thus effecting passivation. These findings disprove "the adsorption theory of passivation" and "the theory of passivity film" and inform a new theory we call "passivation theory of incremental oxidation damping".

## Main Text:

**P**assivation produces passive metals and alloys to underpin modern civilization. While recognized as one of the most important phenomena (*1, 2*), it has been defying full scientific explanation for more than 170 years. The theory of passivity film and the adsorption theory of passivation, equally powerful, are pulling us toward totally different direction. On the surface, they agree that an underlying chemisorbed monolayer is formed first and an upper oxide layer is formed later, and the two layers compose a passive film. But they divide on the role played by the layers. The former asserts that this dense, tightly attached passive layers with an independent phase serves only as a physical barrier to separate metal matrix from the environment. Evans reported successful separation of passive film from Fe-Ni steel (*3-5*). But the latter tells another story: the formation of a stable two-dimensional (2D) chemisorbed monolayer of O or $O^{2-}$ containing particles ($O^{2-}$ or $OH^{-}$) on the metal surface or part of the surface is enough to reduce the reactivity of the outmost metallic atoms, while the subsequent formation of oxide layer above the chemisorbed monolayer only plays no more than a subordinate role (*6*). Evidence is from electron diffraction studies that the 2×2 adsorbed oxygen structure formed on Ni surface is much more stable than NiO.

These two standard theories reveal as much as they conceal. They take overly-complex passive models and adopt defective electrochemical characterization. The most widely investigated passive films are from alloys which are prepared through electrochemical methods. The passive films so prepared include two or more metallic elements with different properties, processed under complex external conditions and multiple actions of numerous reactive oxygen species (ROS) (*7*) and composed of more than complex oxide products (*8, 9*). For this very complexity, the intrinsic structure feature of the passive film is simply out of intellectual grasp. To make matter worse, the electrochemical analysis is a secondary analysis technique based on electrical signals (*10, 11*), incapable to capture the fine structure of the passive film.

The popular theoretical approaches should be blamed. The efforts to study kinetic behaviors and mechanism of passivation and the structure of a passive film were focusing on microscopical and even atomic levels (*12, 13*), with the microscopical disorder being embraced as a basic structural feature of all passive films (*14-16*). This cannot be true. The formation of ultrathin passive film is the emergence of a complex macroscopic system. Its acroscopic structural features, from the emergence theory (*17, 18*), might not be represented or revealed by the microscopical features and laws. Moreover, numerous attempts to explore dynamics mechanism of the passivation are directed at properties of thermodynamic equilibrium systems. But metallic passivation is indeed far from thermodynamically stable state (*19, 20*). In many cases, metals are oxidized to thermodynamically stable oxides in air. This causes peeling of the oxides and degeneration of metals, leading to failure of various applications (*21*) which account for trillions every year globally (*22*). In stark contrast, consider a simple Ti passive film in air. Passivation inputs environmental oxygen into Ti metal, oxidizing it into two subsystems, a passive film and the internal metallic matrix. This passive film tightly attached to the internal metallic matrix is far removed from equilibrium state but enabled functionally to protect the matrix from oxidation corrosion. It can even remold Ti properties (*23-26*).

We conceive passivation as the emergence of a macroscopically ordered subsystems. We design totally new experiments to investigate passivation of Ti and other like metals. We arrange a set of three kinds of gradient-upgrading oxidizing stimuli to treat fresh Ti discs at room temperature: NA-Ti group, treated under the atmospheric environment; $O_3$-Ti group, treated under the atmospheric environment which contains pumped $O_3$ (42.5±5.3mg/h); O-Ti group, treated under UV (188nm) irradiated atmospheric environment that includes UV-split O atoms and $O_3$ (44.4±3.9mg/h) formed via the combination of O atoms and $O_2$. By comparing the results, we find that passivation leads to the emergence of a spatiotemporally ordered passive membrane, rather than a passive film. The membrane is composed of the thermodynamically stable Ti-O chemisorbed monolayer and the underlying gradient oxide layer far removed from equilibrium state. This ordered structure actually "grows" to effect passivation, in the sense that it comes from a nonlinear mechanism of incremental oxidation damping. For like metal, zirconium (Zr), niobium (Nb), tantalum (Ta), nickel (Ni) and even the main group element aluminum (Al), we find the same. These findings can be explained neither by the theory of passivity membrane nor by the adsorption theory. Rather, they call upon us to come up with a new theory for passivation.

## Results

### Ti passive membrane grows in gradient order

Limited by the electron escape depth, the detection depth on Ti surface of XPS is about 2~3nm, the upper of the passive membrane. The three experimental groups exhibit a gradient structural feature in Ti oxidation degree. With increasing passivation time, the oxidation degree grows in gradient order. Meanwhile, the oxidation degree of passive membranes is positively correlated with the oxidizing ability of the environmental oxygen species within the same passivation time span. The ratio of $Ti^{metallic}/Ti^{4+}$ is used to stand for oxidability of Ti passive membrane. For a fresh Ti disc, in NA-Ti groups the ratio of $Ti^{metallic}/Ti^{4+}$ decreases by 3.7% within 96 hours, while 9.1% in $O_3$-Ti and 12.0% in O-Ti (Fig. 1A and fig. S1). The oxidation degree shows the highest rise in O-Ti group, followed by $O_3$-Ti and NA-Ti. Even so, $Ti^{metallic}$ peaks assigned to Ti matrix still exist in the O-Ti group. As expected, after ageing for one year, the oxidation degree of NA-Ti increases further (fig. S2). Interestingly, the oxidation degree of Ti disc aged for one year is even lower than O-Ti aged for 96 hours. This shows Ti passive membrane grows much faster in O atom-involved air than in air. The results of Angle-resolved XPS measurement, as shown in Fig. 1B and fig. S3A-B, also reveals the structural feature of gradient descent in oxidation degree for Ti passive membranes.

The XPS spectra of Ti*2p* at varying depths of passive membranes on NA-Ti, $O_3$-Ti and O-Ti show that the three gradient-upgrading oxidizing stimuli create three passive membranes of increasing oxidation degree. Each passive membrane shows a gradient structural feature of oxidation degree. And each depth contains four valence states of Ti (Fig. 1C), however, they differ in content. At the upper of the three membranes, $TiO_2$ component takes up a 70-80% Ti*2p* signal and is the highest in O-Ti. As the depth increases, the fractions of suboxides and metallic Ti increases rapidly, and suboxides is predominant from 2nm to 8nm. The least $TiO_2$ is detected at 8nm of the NA-Ti, following by $O_3$-Ti and O-Ti. Regardless of the oxidation treatment, the fraction of Ti is the majority at the bottom of the passive membranes.

The EDS mapping analysis presents a O-rich layer. With increasing oxidation stimuli, the passive membranes grow thickened, and the atomic amount ratio of O/Ti increases from 1.29 of NA-Ti to 1.45 of $O_3$-Ti and 1.56 of O-Ti (Fig. 1D). Represented as line profiles across the passive membranes, the components of Ti and O demonstrates opposite gradient trends in the three passive membranes (Fig. 1E). Clearly, Ti passive membrane grows in gradient order.

### The outmost layer is a thermodynamically stable chemisorbed Ti-O monolayer

UPS detection information is more surface sensitive (~10Å, the upper of the oxide layer). After 96 hours' treatment by $O_3$ or O atoms, values of work function ($\varphi$) are 4.67eV and 4.74eV respectively, higher than 4.36eV of NA-Ti and 4.33eV of cp Ti, while much lower than $\varphi$ of $TiO_2$ 5.58eV (Fig. 2A). While stronger oxidation by $O_3$ or O atoms enhances the valence electron stability of the passive membrane and reduces their electron transfer and oxidation rates, the upper of the two passive membranes are still apt to be further oxidized. This is consistent with the angle-resolved XPS results that the top layer of the oxide layer still contains Ti matrix and suboxides (Fig. 1B). This suggests that the passive membrane is covered with a thermodynamically stable structure which is neither Ti oxides nor Ti metal, while it is stable enough even in the O atoms contained environment and can prevent the direct action of O atoms on the underlying suboxides and metallic matrix. If not, assuming the outmost of the passive membrane is suboxides and the residue metallic matrix, O atoms should rapidly oxidize them into $TiO_2$ structure that is unable to be further oxidized.

According to the 9 crystal planes present in the XRD detection (Fig. 2B), calculated O atom adsorption stability is further shown in Fig. 2C. Of the 9 crystal planes, the plane (200) has the lowest content (~1%). In the view of energy, except the plane (200), the chemisorbed Ti-O monolayer is more stable than TiO in the other 8 planes. These results tell that the outmost layer of the passive membrane is this thermodynamically stable Ti-O chemisorbed monolayer. The reason lies in that under our experimental conditions, once the chemisorbed Ti-O monolayer forms, in terms of kinetics and thermodynamics of reactions, it is unlikely for the underlying Ti atoms to escape from the lattice structure and cross this monolayer to react with the environmental oxygens. Instead, it is easy for the environmental oxygens of high energy to be absorbed on this monolayer and pass through it to react the Ti matrix underneath further forming the oxide layer. This contradicts the current two passivation theories which, although lack of evidence, hold that the chemisorbed monolayer is under the oxide layer (*27*).

**The gradient oxide layer under the Ti-O monolayer is far removed from equilibrium**

Aberration-Corrected TEM (ACTEM) reveals the thickness of passive membrane is 7.1±0.2nm of O-Ti and 6.3±0.2nm of $O_3$-Ti, thicker than 5.3±0.5nm of NA-Ti [Fig. 3A(a), Fig. 3B(a) and Fig. 3C(a)]. Although a thin membrane can be seen to cover around Ti matrix at low magnification, there is no definite boundary between the membrane and Ti matrix at atom scale. From the upper of the oxide layer to the bottom (fig. S3C- fig. S4), with the amorphous oxides gradually withering, the Ti lattice gradually increase until reverting to Ti matrix. The amount and area of lattice structures decrease from the bottom to the surface, while those of the amorphous structures increase. After $O_3$ or O atoms treatment, the lattice structures become smaller but more while the amorphous structures get larger. Nevertheless, in O-Ti group, the nanoscale lattice structures still exist in upper of the passive membrane.

Interestingly, fast Fourier transform (FFT) pattern of any region in the three passive membranes is almost the same as the lattice structure of Ti matrix [Fig. 3A(b), Fig. 3B(b) and Fig. 3C(b)]. Together with the $Ti^{metallic}$ existence detected by XPS, these nanoscale lattice structures are attributed to residue Ti lattice structure. In a room temperature atmosphere, subject to high energy barrier of the formation of oxide crystals and stability of the Ti lattice structure, the minimum energy principle determines that the passivation in the atmosphere at room temperature is more likely to leave the residue of Ti lattice structures rather than the formation of $TiO_2$ nanocrystals in Ti passive membrane. This is different some reports that the nanocrystals in electrochemical passivation are oxide (*28-30*). The reason may lie in that the electrochemical passivation has a much stronger driving force, which provides energy enough to produce oxide nanocrystals in the passive membranes. Even so, the electrochemically-made passive membranes are still far removed from equilibrium, for metallic components and suboxides still exist in them (*31-34*).

Ti L2- and L3-edge EELS spectra shows that the centers of the L3 and L2 peaks are chemically shifted to higher energy losses as the oxidation degree increases (*35, 36*). As shown in Fig. 3A(c), Fig. 3B(c) and Fig. 3C(c), four typical positions are selected, the outmost, the middle, the bottom of the passive membranes and the Ti matrix adjacent to the passive membranes. Disappearance of the O signals at the Ti matrix adjacent to the passive membrane exhibits a definite existence of an oxide layer. From the outmost to the bottom, Ti L3 and L2 peaks are chemically shifted from higher energy losses to lower energy losses, indicating a gradient decline of Ti oxidation degree. Moreover, compared to NA-Ti group, the L3, L2 peaks of the middle and the bottom shift towards those of the outmost layer in the O-Ti group or $O_3$-Ti group, and the middle peaks almost coincide with the outmost. Clearly, after the treatment by $O_3$ or O atoms, Ti atoms at the middle layer shows a more similar chemical environment to those at the outmost layer. On the other hand, although XPS results exhibit the highest contents of $Ti^{4+}$ in the three outmost layers, the characteristic four-peak of EELS spectra of $Ti^{4+}$ is not found. This indicates no aggregate structure of $TiO_2$ in the passive membranes. Therefore, the nanoscale lattice structures in Ti passive membrane are the residue Ti lattice structures, and the oxide layers are of a TiOx@Ti cermet-like gradient structure.

**Structural features and fluctuation of passive membranes of other passive metals**

Through like experiments, the structural features and fluctuations of passive membranes of Zr, Nb, Ta, Ni and Al formed in air have been investigated. Surface sensitive UPS detection reveals that even in O atom-containing environment, continuous oxidation behaviors also are found for all these passive membranes, just like that of Ti (Fig. 4A, fig. S5 and fig. S6). Meanwhile, all the passive membranes are composed of residue metallic matrix and oxides, and exhibit a gradient characteristic of component distribution and oxidation degree (Fig. 4B, fig. S7-fig. S8). And their structural fluctuations with the oxidative stimuli are also similar to that of Ti passive membrane.

## Discussion

**Emergence of Ti passive membrane breaks symmetry in the direction of oxidation**

This set of structural fluctuation experiments shows that passivation transforms a locally-ordered but macroscopically-disordered Ti surface into a spatiotemporally-ordered passive membrane. Ti passive membrane is composed of the thermodynamically stable Ti-O chemisorbed monolayer and the underlying gradient oxide layer far removed from equilibrium state. More important, between the passive membrane and the internal metallic matrix, there exists a nonlinear interaction mechanism, by which under stronger oxidative stimuli Ti passive membrane grows into a thicker one with higher degree of oxidation to resist the external environmental stimuli, so as to better protect the inner Ti matrix from oxidation.

The emergence of a spatiotemporally ordered subsystem indicates a decreased entropy of the open system (Fig. 5A). This upends our current understanding of passivation. Intuitively, the passivation seems to destroy the local orderliness of upper microstructure of crystal grains in Ti surface and increases the entropy of the open system, and numerous reports also implied a micro-disordered structural feature of a passive film. Indeed, the formation of a thermodynamically stable Ti-O chemisorbed monolayer indicates the Ti outmost atoms form a continuous 2-D ordered structure ($S1-S0<0$). And the passivation entraps much disordered environmental oxygens into the metallic system spontaneously producing a gradient spatial distribution of O from the upper to the bottom of Ti passive membrane ($d_eS1<0$). Accordingly, Ti is compressed into a reverse gradient spatial distribution ($d_iS1<0$). In addition, the average oxidation degree of Ti ($d_iS2<0$), the amorphous oxides ($d_eS2<0$) and the residue Ti lattice structures ($d_iS3<0$) also exhibit a gradient feature. The emerged orderliness in these dimensions also reveals the broken symmetry in the direction of oxidation.

From the perspective of energy and dynamics, together with capturing disordered gas oxygen species in the environment and absorbing their kinetic energy, the exothermic passivation causes the evolvement from a simple local-ordered but macro-disordered surface of polycrystalline Ti into a spatiotemporally-ordered macroscopical subsystem of passive membrane. In particular, the O/Ti ratio is much above 1, showing that a huge negative entropy is introduced in the open system. Meanwhile, the gradient distribution of Ti oxidation degree, the amorphous oxides and the residue lattice structures tell an ordered distribution of both electron cloud and newly-born localized structures in the resultant passive membrane. This leads to a net decrease in the whole entropy of the open system ($\Delta S = d_iS1 + d_iS2 + d_iS3 + d_iS4 + d_eS1 + d_eS2 < 0$) (Fig. 5B). The emergence of an ordered but complex macroscopical subsystem enables Ti to evolve numerous macro-properties such as the high stability, antioxidant capacity, corrosion resistance, and even excellent gas gettering abilities (*37, 38*) and biological properties.

**A non-linear kinetics behavior and mechanism of Ti passivation**

The broken symmetry of structure of Ti passive membrane in the direction of oxidation is satisfactorily explained by a non-linear kinetic behavior of passivation. As one of transition metals with the D-electron shell of Ti not being completely filled, the outmost atoms of Ti are prone to induce ultrafast chemical adsorption of oxygen on these metal surface at room temperature. The formation of the thermodynamically stable chemisorbed Ti-O atom monolayer immediately inactivates the outmost Ti atoms of metal, and also shield the direct reaction of various environmental stimuli such as $O_2$, $O_3$ and O atoms on the metallic matrix.

Subsequently, numerous environmental oxygens are adsorbed onto this chemisorbed monolayer and are polarized into O atoms. Under the induction of the active electrons of Ti atoms underlying the chemisorbed monolayer, the O atoms pass through the monolayer and are converted into $O^{2-}$ ions indiscriminately. Rapid accumulation of $O^{2-}$ ions immediately results in a very high oxidation potential under the monolayer. This potential in turn promptly drives $O^{2-}$ ions to diffuse toward the metal interior through the physical barrier of Ti lattice structure in a diminishing way. This crashes the surface Ti lattice structure randomly and preferentially produce various oxides adjacent to the cracks, inducing the emergence of a TiOx@Ti ceramet-like oxide layer with a gradient structure. Immediately after the formation of a passive membrane, the combined effect of the chemical inactivation, the physical barrier and inverse potential of the membrane in turn resists the oxidation potential. The growth of passive membrane slow sharply and the open system falls into the passive state. Moreover, since O atom or $O_3$ has a much higher reactivity compared to $O_2$, their adsorbed rate and reactive efficiency on the monolayer is much higher. Accumulation of much more $O^{2-}$ ions cause a much higher oxidation potential within the same time span. Therefore, the O-Ti has the thickest passive membrane with the highest oxidation degree, followed by the $O_3$-Ti group and the $O_2$-Ti group.

Upon the formation of passive membrane, between the numerous residue lattice structures and the amorphous oxides, there exists an interface with large specific surface area, where the micro-strain by misfit causes structural defects rich in active sites. This is confirmed by the micro-strain ε value in thin-film X-ray diffractometry (TF-XRD) that there are more micro-strains in the oxide layer after $O_3$ or O atom treatment (O-Ti$|\varepsilon| >$ O3-Ti$|\varepsilon| >$ NA-Ti$|\varepsilon|$) (fig. S9). Like diffusion channels of chloride ions (*29*), the interface regions provide favorable channels for $O^{2-}$ diffusion. Under the continuous stimuli of environmental oxygen species, $O^{2-}$ ions driven by the oxidation potential further cracks the residue lattice structures into smaller structures and increases the oxidation degree of amorphous oxides while invading deeper matrix producing a thicker oxide layer. The stability of the resultant membrane is further increased such that they are able to resist higher oxidation potential (fig. S10, table. S1 and table. S2). Now it is clear that in a passivating environment, the oxidation degree and thickness of a passive membrane increases with time goes on.

For the first time, a non-linear kinetic model of metal passivation is given at the macro level. The ordered structural features and the non-linear kinetics behavior of Ti passive membrane together shows a more convincing and explanative passivation mechanism. This mechanism relies on the synergistic effect of the thermodynamically stable chemisorbed monolayer and the underlying gradient oxide layer far removed from equilibrium. The chemisorbed monolayer is able to convert the environmental oxygens into $O^{2-}$ ions indiscriminately, and the non-equilibrium underlying oxide layer is able to continuously absorbs the diffused $O^{2-}$ ions driven by the oxidation potential. This makes the gradient Ti passive membrane being continuously fortified via breaking symmetry in the direction of metallic oxidation in a passivating environment. Ti thus evolves a potent resistance to various oxidative or corrosive environments.

We name this non-linear passivation mechanism as the "passivation mechanism of incremental oxidation damping" (Fig. 5C). This passivation mechanism is dissipative. Different from the classic dissipative patterns such as Benard convection, laser and chemical oscillation (*39-41*), relying on this non-linear passivation mechanism, an ordered solid-state passive membrane keeps growing slowly to dissipate the continuous oxidation stimuli of the environmental oxygens on Ti matrix thus effecting passivation.

**Extrapolation of the macro-ordered structural feature of Ti passive membrane**

Our findings suggests that metallic passivation comes from the emergence and growth of a spatiotemporally ordered passive membrane. Passivation in ambient air is one of the most common forms of metallic passivation, and the inherent structural features of the resulting simple passive membrane and the passivation mechanism of dissipation are found elsewhere. Ni, as one of the first transition series metals like Ti and also a basic experiment model for the adsorption theory of passivation, follows the same passivation law. As are further verified by the structure features of the passive membranes of Zr, Nb and Ta, which belongs to the secondary transition series metals and the tertiary transition series metals, respectively. As a main group metal, Al passive membrane also shares similar structural features.

As further evidence, the macro-ordered gradient features of dissipation are also found in the two electrochemically made Ti passive membranes in 1 M $H_2SO_4$ solution or 1 M NaOH solution (fig. S11). Interesting, the gradient-ordered feature is also found to extensively exist in investigations of various electrochemically-made passive membranes (*12, 13, 42-45*). In fact, although the electrochemically made passive membranes are of variety and their component and local structure are extremely complex, all them also share two common features: one is a spatiotemporally-ordered passive membrane consisting of both a thermodynamically stable outmost layer against the external oxidative environments and an underlying gradient non-equilibrium oxide layer; the other is to effect passivation relying on their growth via nonlinear passivation mechanism. This paves fresh ground for grasping metallic passivation.

In particular, when passivating voltage is high to anodizing, resultant double-layer passive membranes seem like totally different from this simple Ti passive membrane. Nevertheless, they also follow the dissipative theory. An outer porous layer and an inner dense layer compose these double-layer passive membranes (*46*). Typically, although the porous outer seems like loose, the porous layer is a continuous seamless thermodynamically-stable spatial closed curved surface. According to this study, the loose structure of the porous outer layer, as the thermodynamically stable oxidative product of the inner gradient dense layer under high anodizing voltage, may serve as the role just as the chemisorbed monolayer of a simple passive membrane. The thermodynamically stable loose layer and the non-equilibrium dense layer work synergistically to keep the double-layer passive membranes growing thus effecting passivation (Unpublished Data). The relationship and difference and transformation between the simple Ti passive membrane and the electrochemically made passive membranes is basically solved in our lab and will be reported in the near future.

**Overcome defects of exiting passivation theories**

*Our findings have cracked the secret spatiotemporal relationship between the chemisorbed monolayer and the oxide layer that long puzzled the passivation field*. The two prevailing passive theories held that the oxide layer subsequently formed was above the chemisorbed monolayer. Although it is not thermodynamically tenable, it was difficult to disprove it for the investigations of passivation in the past were basically based on the electrochemical passivation. The reason lies in that the passivation behaviors and the structure of resultant passive films are too complicated due to the strong driving force of passivation and the too complicated passivation conditions. However, our study demonstrates that it was false. Actually, the prerequisite and necessary for the emergence and growth of a non-equilibrium passive membrane is that its outmost layer is thermodynamically-stable against the external environments, which should be taken as one of the basic principles in building the theory of metallic passivation. This also applies to some extreme cases. For instance, the formation a stable chemisorbed monolayer on the metal surface or part of the surface is sufficient to passivate metal, like electrochemical passivation of Fe or Zn in alkaline solutions (*47, 48*). Assuming without its thermodynamical stability against the external environment stimuli, it is impossible for this type of passive membranes to exist. Let alone growth.

*Our findings disclose the inherent regularity of metallic passivation comes from the spatiotemporally orderliness of the structure and functions of a passive membrane*. In the dimension of time, the successive formation of the chemisorbed monolayer and the oxide layer compose the two stages of passivation. From spatial structure, the chemisorbed monolayer is on the gradient oxide layer. In terms of thermodynamics, the outmost thermodynamically-stable Ti-O monolayer provides the precedent condition for the formation of a passive membrane. And the non-equilibrium oxide layer provides the structural basis for the continuous growth of a passive membrane. Functionally, relying on the mechanism of the incremental oxidation damping, the Ti-O monolayer and the oxide layer work synergistically to keep the passive membrane growing orderly and slowly thus effecting passivation. Both of them play different roles in the passivation respectively, but they are interdependent and inseparable.

This study informs a theory we call “passivation theory of incremental oxidation damping”, that scientifically unifies the formation and roles of the chemisorbed monolayer and the underlying solid membrane into the passivation process. The passive membrane is the product of the interactions between an open system and external environment and also the structural basis of the matter and energy exchange, and is essential for the self-becoming, self-being and self-protection of an open system. This gives us a hint that the formation of a macro-ordered interface of dissipation between an open system and environment may be the condition precedent of heterogeneity of everything in the physical world and also be a general interfacial phenomenon from an inanimate matter to an animate life entity. As a result, our findings and theory might also contribute to general understanding of nature and science.

## References and Notes

1. D. D. Macdonald, Passivity-the key to our metals-based civilization. *Pure Appl. Chem*. **71**, 951-978 (1999). doi: 10.1351/pac199971060951
2. P. Marcus, *Corrosion mechanisms in theory and practice* (Marcel Dekker Inc., New York, 2002).
3. U. R. Evans, The passivity of metals. Part I. The isolation of the protective film. *J. Chem. Soc*. 1020-1040 (1927). doi: 10.1039/jr9270001020
4. U. R. Evans, Isolation of the film responsible for the passivity of an iron anode in acid solution. *Nature*. **126**, 130-131 (1930). doi: 10.1038/126130b0
5. U. R. Evans, Inhibition, passivity and resistance: a review of acceptable mechanisms. *Electrochim. Acta*. **16**, 1825-1840 (1971). doi: 10.1016/0013-4686(71)85141-1
6. H. H. Uhlig, The adsorption theory of passivity and the flade potential. *Z. Elektrochem*. **62**, 626-632 (1958). doi: 10.1002/bbpc.19580620603
7. T. Ohtsuka, A. Nishikata, M. Sakairi, K. Fushimi, *Electrochemistry for corrosion fundamentals* (Springer Singapore, Singapore, 2018).
8. I. Milošev, M. Metikoš-Huković, H. H. Strehblow, Passive film on orthopaedic TiAlV alloy formed in physiological solution investigated by X-ray photoelectron spectroscopy. *Biomaterials*. **21**, 2103-2113 (2000). doi:10.1016/S0142-9612(00)00145-9
9. S. Virtanen, P. Schmuki, H. S. Isaacs, In situ X-ray absorption near edge structure studies of mechanisms of passivity. *Electrochim. Acta*. **47**, 3117-3125 (2002). doi: 10.1016/S0013-4686(02)00230-X
10. ES. M. Sherif, F. H. Latief, H. S. Abdo, N. H. Alharthi, Electrochemical and spectroscopic study on the corrosion of Ti-5Al and Ti-5Al-5Cu in chloride solutions. *Met. Mater. Int.* **25**, 1511-1520 (2019). doi: 10.1007/s12540-019-00306-2
11. X. Gai *et al.*, Electrochemical behavior of passive film formed on the surface of Ti-6Al-4V alloy fabricated by electron beam melting. *Corros. Sci*. **145**, 80-89 (2018). doi: 10.1016/j.corsci.2018.09.010
12. L. Wang *et al.*, Quantitative analysis of local fine structure on diffusion of point defects in passive film on Ti. *Electrochim. Acta*. **314**, 161-172 (2019). doi: 10.1016/j.electacta.2019.05.048
13. L. Wang *et al.*, Local fine structural insight into mechanism of electrochemical passivation of titanium. *Acs. Appl. Mater. Inter*. **8**, 18608-18619 (2016). doi: 10.1021/acsami.6b05080
14. P. Tengvall, I. Lundström, Physico-chemical considerations of titanium as a biomaterial. *Clin. Mater.* **9**, 115–134 (1992). doi: 10.1016/0267-6605(92)90056-Y
15. Y. Yang *et al.*, Liquid-like, self-healing aluminum oxide during deformation at room temperature. *Nano. Lett*. **18**, 2492-2497 (2018). doi: 10.1021/acs.nanolett.8b00068
16. X. X. Wei *et al.*, Enhanced corrosion resistance by engineering crystallography on metals. *Nat. Commun*. **13**, 726 (2022). doi: 10.1038/s41467-022-28368-8
17. J. H. Holland, *Emergence: From Chaos to Order* (Addison-Wesley Longman Publishing Co. lnc., New York, 1998).
18. P. W. Anderson, More is different. *Science*. **177**, 393-396 (1972). doi: 10.1126/science.177.4047.393
19. I. Prigogine, *From Being to Becoming* (W. H. Freeman & Co, San Francisco, 1980).
20. I. Prigogine, Time, Structure, and Fluctuations. *Science*. **201**, 777-785 (1978). doi: 10.1126/science.201.4358.777
21. R. C. Newman, K. Sieradzki, Metallic Corrosion. *Science*. **263**, 1708-1709 (1994). doi: 10.1126/science.263.5154.1708

22. G. Koch, *In Trends in Oil and Gas Corrosion Research and Technologies* (Woodhead Publishing, Boston, 2017), pp. 3-30.
23. B. Kasemo, Biocompatibility of titanium implants-surface science aspects. *J. Prosthet. Dent*. **49**, 832-837 (1983). doi: 10.1016/0022-3913(83)90359-1
24. G. Lütjering, J. C. Williams, *Titanium* (Springer-Verlag, Berlin, ed. 2, 2007).
25. R. Adell *et al.*, A 15-year study of osseointegrated implants in the treatment of the edentulous jaw. *Int. J. Oral. Surg*. **10**, 387-416 (1981). doi: 10.1016/s0300-9785(81)80077-4
26. T. Albrektsson *et al.*, The interface zone of inorganic implants *In vivo*: Titanium implants in bone. *Ann. Biomed. Eng*. **11**, 1-27 (1983). doi: 10.1007/bf02363944
27. H. H. Uhlig, Structure and growth of thin films on metals exposed to oxygen. *Corros. Sci*. **7**, 325-339 (1967). doi: 10.1016/s0010-938x(67)80022-2
28. B. Zhang *et al.*, Passivation of nickel nanoneedles in aqueous solutions. *J. Phys. Chem. C*. **118**, 9073-9077 (2014). doi: 10.1021/jp501825e
29. B. Zhang *et al.*, Unmasking chloride attack on the passive film of metals. *Nat. Commun*. **9**, 2559 (2018). doi: 10.1038/s41467-018-04942-x
30. B. Zhang *et al.*, Quasi-situ observing the growth of native oxide film on the FeCr15Ni15 austenitic alloy by TEM. *Corros. Sci*. **140**, 1-7 (2018). doi: 10.1016/j.corsci.2018.06.029
31. Z. Feng *et al.*, Passivity of 316L stainless steel in borate buffer solution studied by Mott-Schottky analysis, atomic absorption spectrometry and X-ray photoelectron spectroscopy. *Corros. Sci*. **52**, 3646-3653 (2010). doi: 10.1016/j.corsci.2010.07.013
32. A. Machet *et al.*, XPS and STM study of the growth and structure of passive films in high temperature water on a nickel-base alloy. *Electrochim. Acta*. **49**, 3957-3946 (2004). doi: 10.1016/j.electacta.2004.04.032
33. H. Luo *et al.*, Passivation and electrochemical behavior of 316L stainless steel in chlorinated simulated concrete pore solution. *Appl. Surf. Sci*. **400**, 38-48 (2017). doi: 10.1016/j.apsusc.2016.12.180
34. K. F. Quiambao *et al.*, Passivation of a corrosion resistant high entropy alloy in non-oxidizing sulfate solutions. *Acta. Mater*. **164**, 362-376 (2019). doi: 10.1016/j.actamat.2018.10.026
35. L. F. Kourkoutis *et al.*, Atomic-resolution spectroscopic imaging of oxide interfaces. *Philos. Mag*. **90**, 4731-4749 (2010). doi: 10.1080/14786435.2010.518983
36. J. S. Kim *et al.*, Nanoscale bonding between human bone and titanium surfaces: osseohybridization. *Biomed. Res. Int*. **2015**, 960410 (2015). doi: 10.1155/2015/960410
37. R. L. Stow, Titanium as a gettering material. *Nature*. **184**, 542-543 (1959). doi: 10.1038/184542a0
38. L. Holland, Sorption of activated gases by titanium films. *Nature*. **185**, 911-912 (1960). doi: 10.1038/185911b0
39. J. Ward. III. William, H. Le. Blanc. JR. Oliver, Rayleigh-benard convection in an electrochemical redox cell. *Science*. **225**, 1471-1473 (1984). doi: 10.1126/science.225.4669.1471
40. S. Ramakrishna, T. Seideman, Dissipative dynamics of laser induced nonadiabatic molecular alignment. *J. Chem. Phys*. **124**, 034101 (2006). doi: 10.1063/1.2130708
41. F. Deng, J. Feng, T. Ding, Chemoplasmonic oscillation: A chemomechanical energy transducer. *ACS. Appl. Mater. Inter*. **11**, 42580-42585 (2019). doi: 10.1021/acsami.9b13723
42. L. Wang *et al.*, In-situ XAFS and SERS study of self-healing of passive film on Ti in Hank's physiological solution. *Appl. Surf. Sci*. **496**, 143657 (2019). doi: 10.1016/j.apsusc.2019.143657

43. R. Jiang *et al.*, Effect of time on the characteristics of passive film formed on stainless steel. *Appl. Surf. Sci*. **412**, 214-222 (2017). doi: 10.1016/j.apsusc.2017.03.155
44. J. S. Noh *et al.*, Effects of nitric acid passivation on the pitting resistance of 316 stainless steel. *Corros. Sci*. **42**, 2069-2084 (2000). doi: 10.1016/S0010-938x(00)00052-4
45. T. Nickchi, A. Alfantazi, Kinetics of passive film growth on Alloy 800 in the presence of hydrogen peroxide. *Electrochim. Acta*. **58**, 743-749 (2011). doi: 10.1016/j.electacta.2011.10.029
46. D. Q. Martins *et al.*, Effects of Zr content on microstructure and corrosion resistance of Ti-30Nb-Zr casting alloys for biomedical applications. *Electrochim. Acta*. **53**, 2809-2817 (2008). doi: 10.1016/j.electacta.2007.10.060
47. B. N. Kabanov, D. I. Leikis, Dissolution and passivation of iron in alkali solutions [in Russian]. *Reports of the Academy of Sciences, USSR*. **58**, 1685-1688 (1947).
48. T. I. Popova, V. S. Bagotskii, B. N. Kabanov, Anode passivation of zinc in alkaline solutions [in Russian]. *Reports of the Academy of Sciences, USSR*. **132**, 639-642 (1960).
49. Coincidentally, the structure and growth mechanism of Ti passive membrane uncovered in this study provides a solid scientific proof for understanding of the acquisition of ultra-high vacuum by using Ti as gettering materials (*37*). It is well known that the gettering properties of the evaporated Ti membrane have a clear dependence on the ionization (atomization) of the residual gas by a Penning gauge. However, the mechanism has been controversial. Previous reports held that “there was no evidence that passivation films on metals grew faster in the presence of active oxygen, but penetration of a metal film by gas atoms so that the crystallites were oxidized on many sides would explain this phenomenon…” (*38*). This set of experiments provides a direct counterexample to his proposal.
50. J. P. Perdew, K. Burke, M. Ernzerhof, Generalized gradient approximation made simple. *Phys. Rev. Lett*. **77**, 3865-3868 (1996). doi: 10.1103/physrevlett.77.3865
51. M. J. Harrison, D. P. Woodruff, J. Robinson, Density functional theory investigation of the structure of $SO_2$ and $SO_3$ on Cu (111) and Ni (111). *Surf. Sci.* **600**, 1827-1836 (2006). doi: 10.1016/j.susc.2006.02.020
52. L. Ren *et al.*, DFT studies of adsorption properties and bond strengths of $H_2S$, HCN and $NH_3$ on Fe (100). *Appl. Surf. Sci*. **500**, (2020). doi: 10.1016/j.apsusc.2019.144232

**Acknowledgments:** We are very grateful to Prof. Huaiwen He (Guanghua Law School, Zhejiang University) for his repeatedly revising and polishing our article, and we benefited a lot from in-depth discussion with him on philosophy and scientific logic. We also gratefully acknowledge for Prof. Yong Wang (Department of Materials Science and Engineering, Zhejiang University), Prof. Hanying Li (Department of Polymer Science and Engineering, Zhejiang University), Prof. Yanwu Xie (Department of Physics, Zhejiang University), and Prof. Yu Kang (Department of Pharmacy, Zhejiang University) for their support and useful discussions. Thank Senior engineer Ya Wang and for her help in FIB samples preparation. Thank Dr. Tulai Sun and engineer Guoqing Zhu for their help in TEM characterization and EDX analyzing.

**Funding:** This work was funded by the National key R&D program of China (2018YFC1105302 and 2018YFC1105304).

**Author contributions:**
Conceptualization: XL
Methodology: XL, WL, QL, YH, XW
Investigation: XL, WL, QL, HX, CL

Visualization: XL, WL, YH, HZ
Funding acquisition: XL, XD
Project administration: XL, WL, XD
Supervision: XL
Writing – original draft: XL, WL, QL, YH
Writing – review & editing: XL

**Competing interests:** Authors declare that they have no competing interests.

**Data and materials availability:** All data are available in the main text or the supplementary materials.

**Supplementary Materials**

Materials and Methods

Figs. S1 to S12

Tables S1 to S2

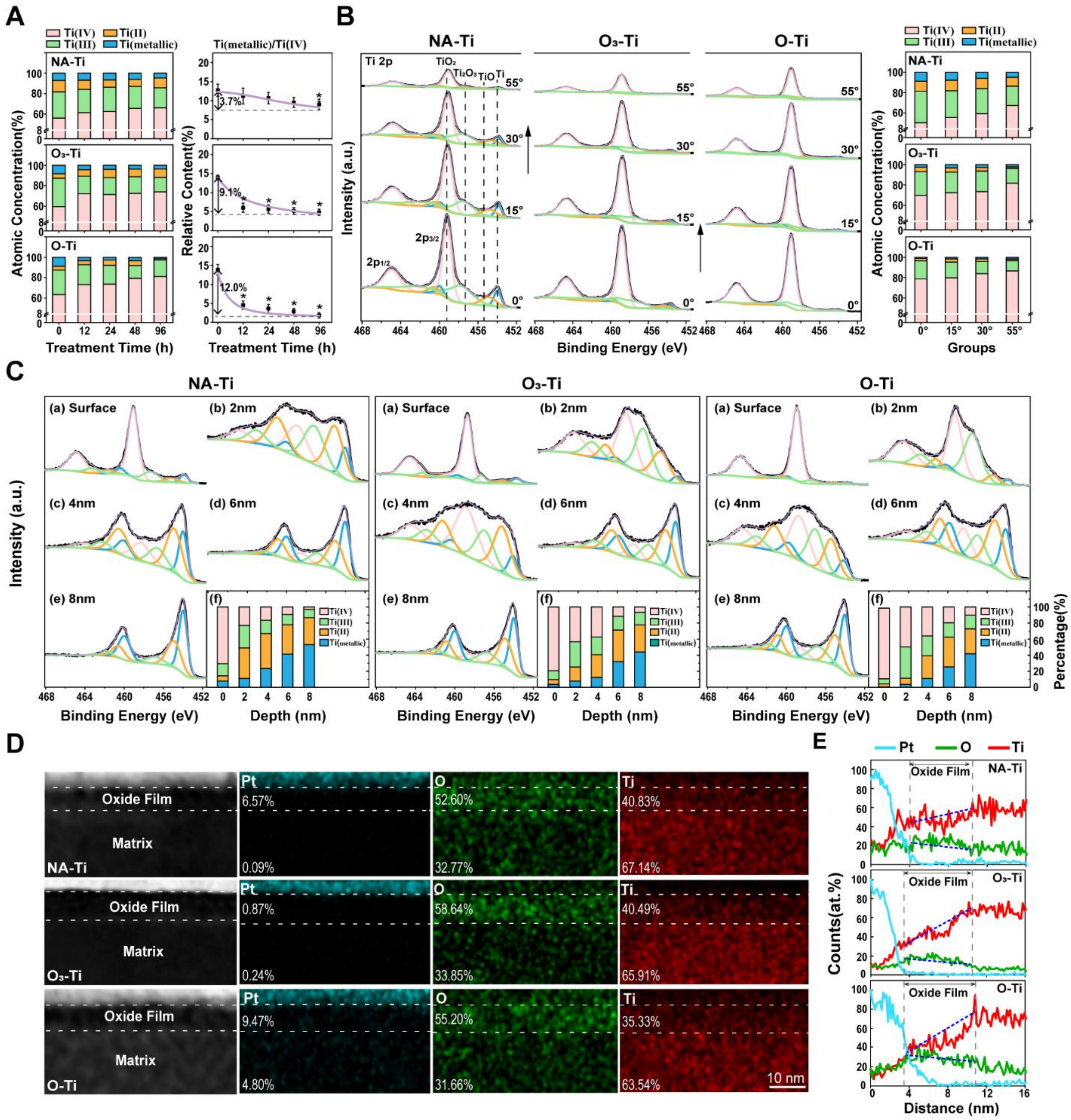

**Fig. 1. Ti passive membrane grows in gradient order.** (**A**) Elemental analysis of XPS spectra on surfaces of NA-Ti, $O_3$-Ti and O-Ti. Left: content percentage of Ti(IV), Ti(III), Ti(II) and Ti(metallic); Right: trend charts of $Ti^{metallic}/Ti^{4+}$ relative content. *$p < 0.05$ vs. treatment time of 0h in the same group. (**B**) Angle-resolved XPS spectra of Ti2p on NA-Ti, $O_3$-Ti and O-Ti with treatment for 48h. Left: Deconvoluted Ti2p peaks in XPS spectra; Right: content percentage of Ti(IV), Ti(III), Ti(II) and Ti(metallic) at different take off angles. (**C**) Deconvoluted Ti2p XPS spectra of (a) surface, (b) 2nm, (c) 4nm, (d) 6nm and (e) 8nm of the passive membranes formed on titanium and (f) percentages of the component peaks to the total intensity of Ti2p at different depths of NA-Ti, $O_3$-Ti and O-Ti, respectively. (**D**) Super-X EDS mapping and (**E**) line profiles of Ti passive membranes on NA-Ti, $O_3$-Ti and O-Ti with treatment for 48h.

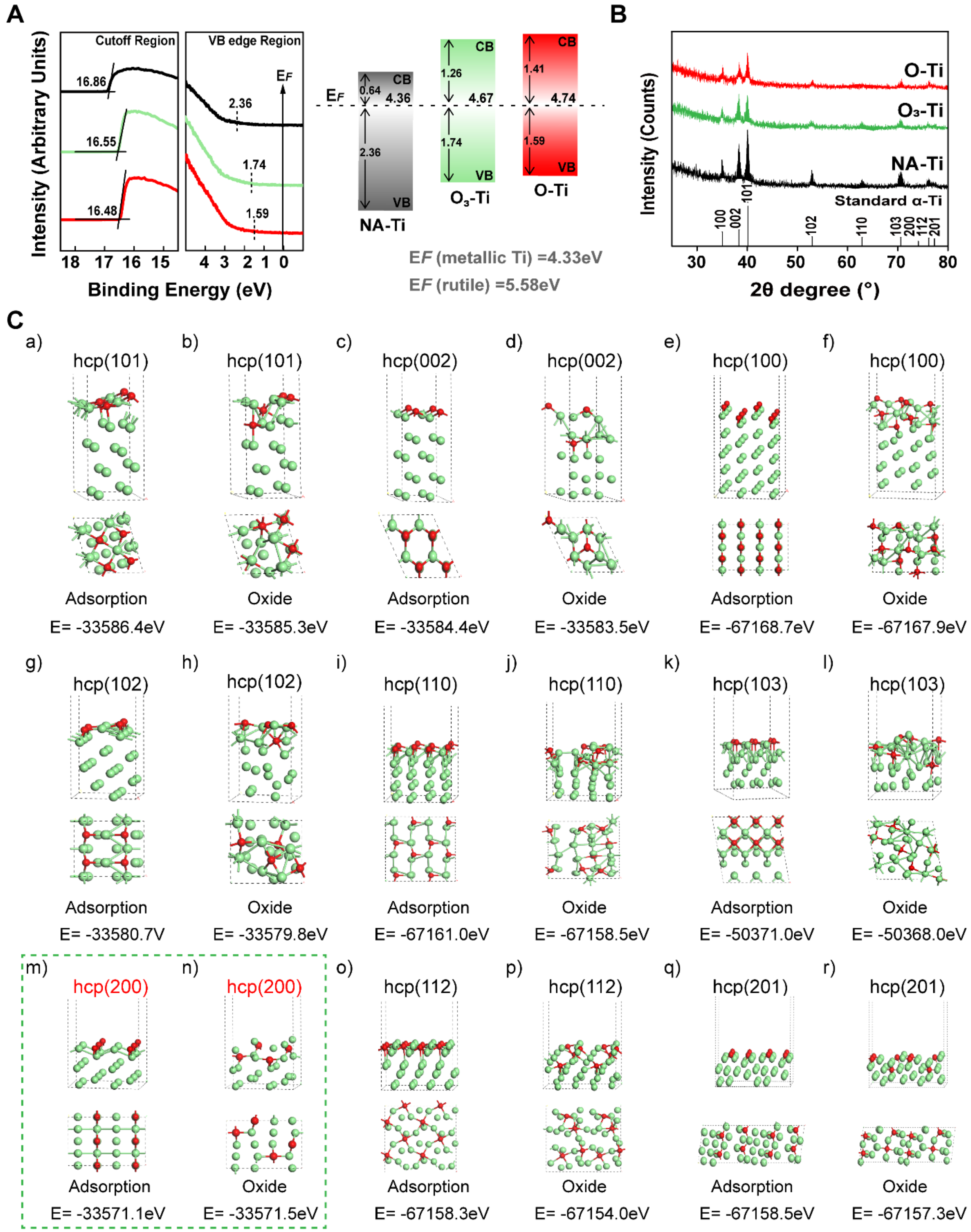

**Fig. 2. The outmost layer of Ti passive membrane is a thermodynamically stable chemisorbed Ti-O monolayer.** (**A**) UPS spectra of the cutoff region and VB edge region for different Ti passive membranes. Spectra from bottom to top represent passive membranes of NA-Ti, $O_3$-Ti and O-Ti with treatment for 96h. Right three columns represent correlative UPS

measured energetic levels. $E_F$, CB, and VB represent Fermi level, conductive band and valence band, respectively. (**B**) TF-XRD results of NA-Ti, $O_3$-Ti and O-Ti with treatment for 48h. (**C**) Side view and top view of adsorption models of O atoms on hcp Ti (101) surface in a), (002) surface in c), (100) surface in e), (102) surface in g), (110) surface in i), (103) surface in k), (200) surface in m), (112) surface in o) and (201) surface in q). Side view and top view of Ti-O oxide models on hcp Ti (101) surface in b), (002) surface in d), (100) surface in f), (102) surface in h), (110) surface in j), (103) surface in l), (200) surface in n), (112) surface in p) and (201) surface in r). The simulated system has better stability with larger absolute value of energy (E). Take the energy at infinity from the nucleus as zero.

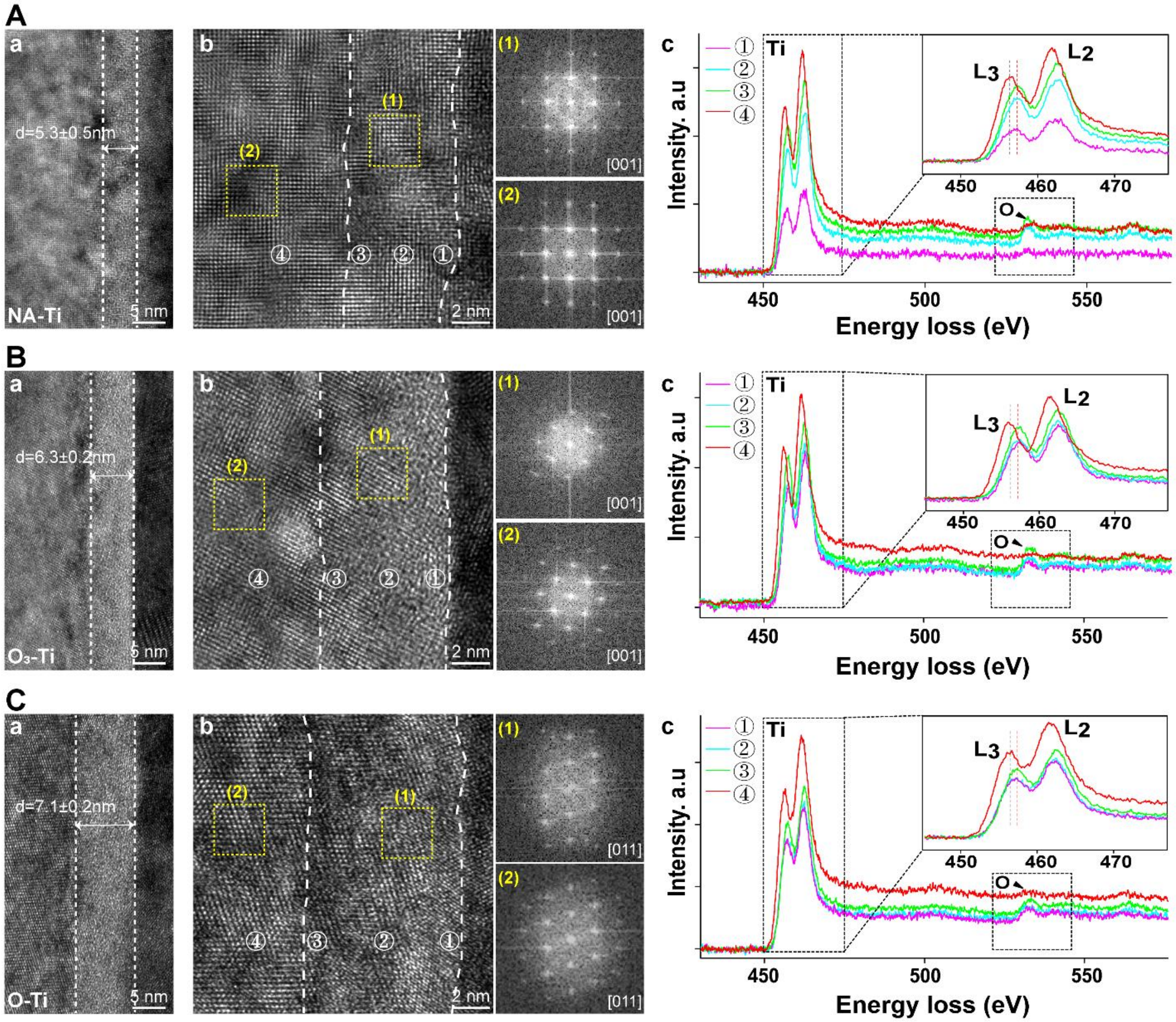

**Fig. 3. Cross-sectional HRTEM images and EELS analysis of the oxide layer of the three Ti passive membranes.** The passive membranes on O-Ti and $O_3$-Ti are thicker than that on NA-Ti with treatment for 48h [**A**(a), **B**(a) and **C**(a)]. Fast Fourier transform (FFT) patterns of selected regions in the passive membrane (1) are almost the same as that of Ti matrix (2) in each group [**A**(b), **B**(b) and **C**(b)]. EELS analysis is performed on NA-Ti, $O_3$-Ti and O-Ti with treatment for 48h [**A**(c), **B**(c) and **C**(c)]. Numbers in left HAADF-STEM image: ①-the upper of the passive membrane; ②-the middle of the passive membrane; ③-the bottom of the passive membrane; ④-the metal matrix adjacent to the passive membrane. There are two characteristic peaks of Ti signals and significant O signals (main adsorbed species in our experiments; Ti L2,3-edge: 462eV&456eV and O K-edge: 532eV) in three different EELS results presented. The centers of the L3 and L2 peaks are chemically shifted to higher energy losses as the degree of oxidation increased.

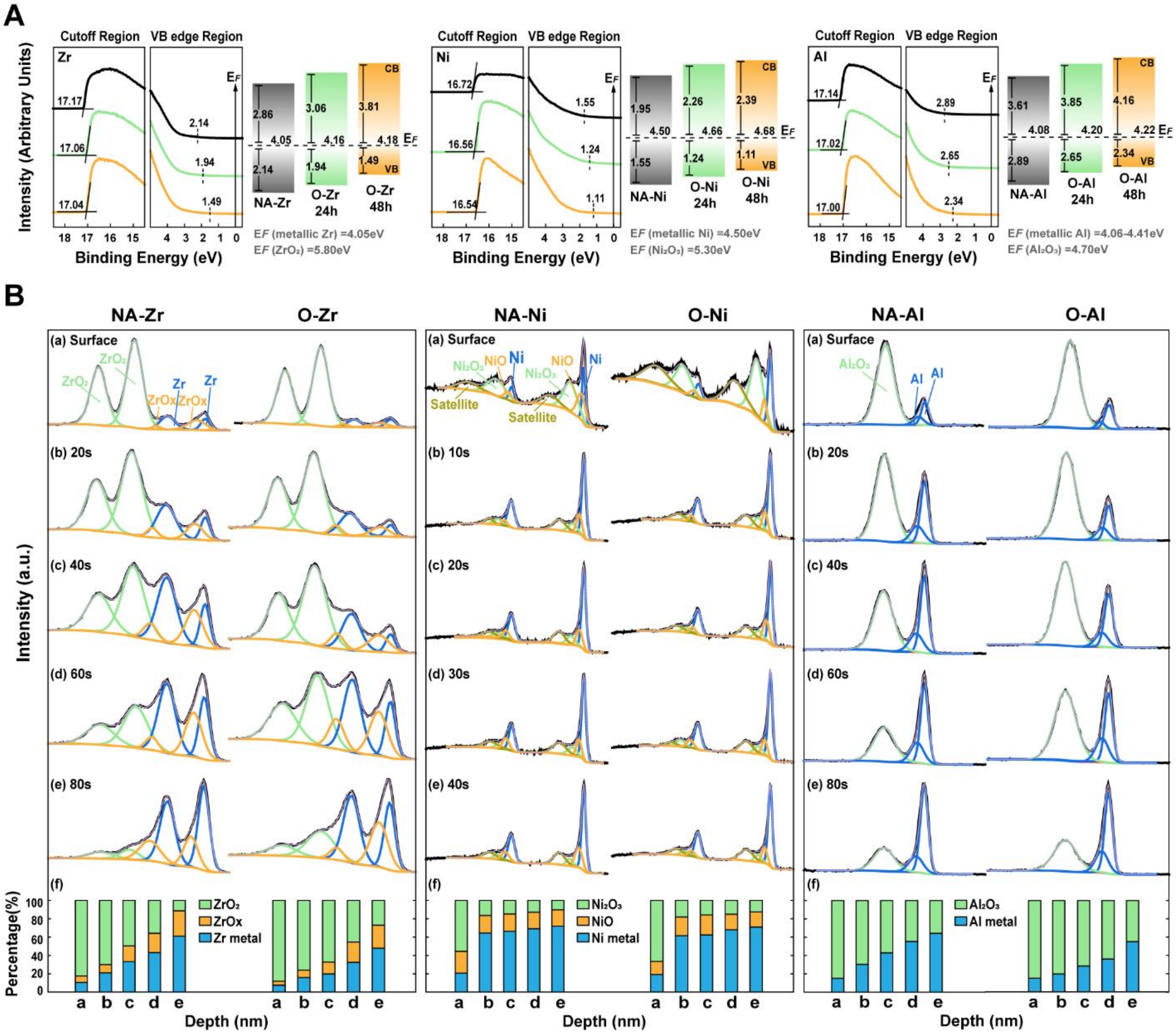

**Fig. 4. Structure features and growth of passive membranes of Zr, Ni and Al.** (**A**) UPS spectra of the cutoff region and VB edge region for Zr, Ni and Al passive membranes. For each passive metal, spectra from bottom to top represent passive membranes of NA-metal, O-metal with treatment for 24h and 48h. Right three columns represent correlative UPS measured energetic levels. $E_F$, CB, and VB represent Fermi level, conductive band and valence band, respectively. (**B**) Deconvoluted Zr3d, Ni2p and Al2p XPS spectra of different sputter times (a) ~(e) of passive membranes formed on Zr, Ni and Al metals and (f) percentages of the component peaks to the total intensity at different depths of NA-metal and O-metal, respectively.

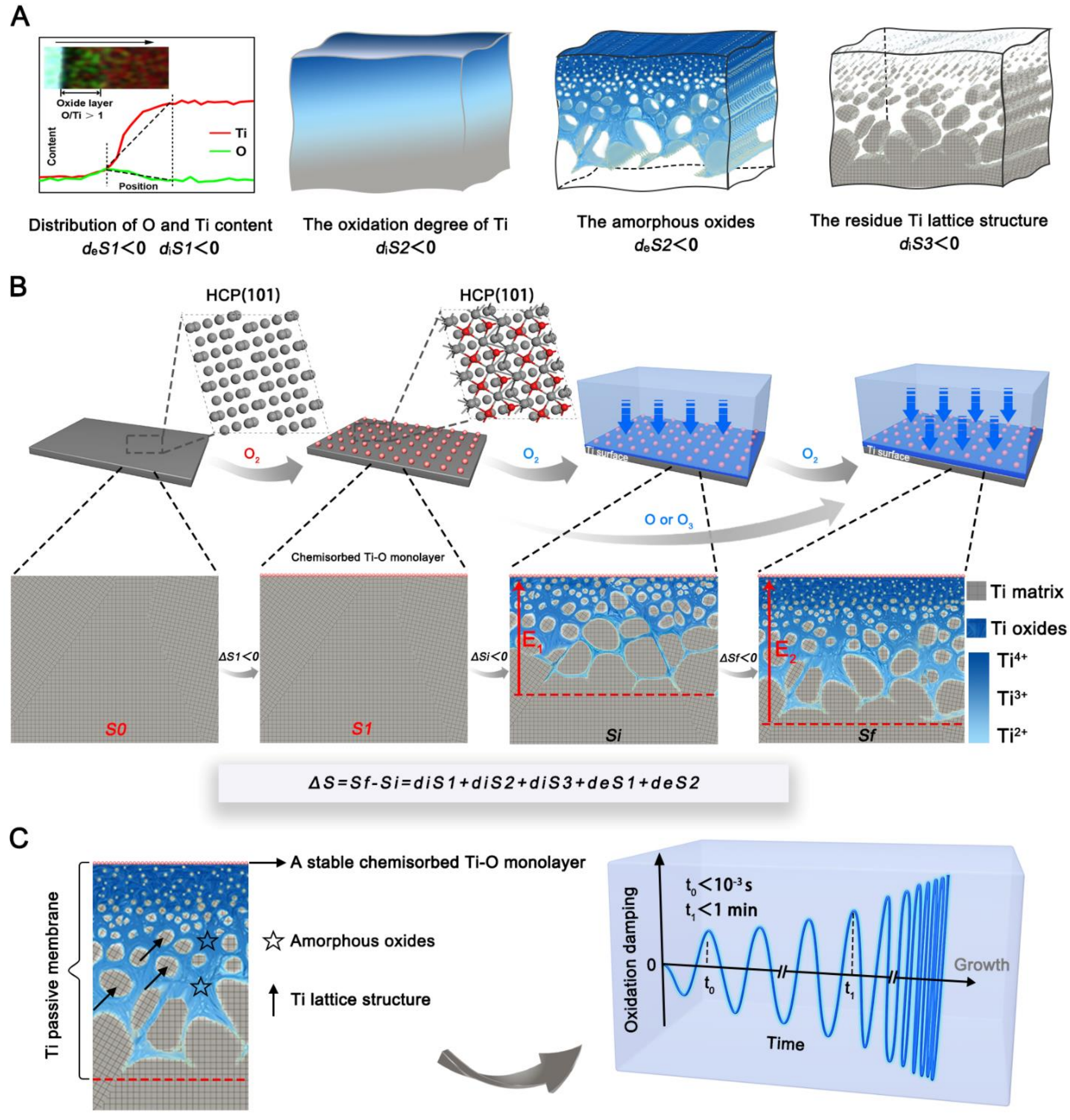

**Fig. 5. Schematic illustration of the formation and growth of Ti passive membrane breaks symmetry in the direction of oxidation.** (**A**) Illustrated details of evolution of system entropy during passivation ($d_eS1$~$S2$<0, $d_iS1$~$S3$<0). (**B**) The formation of spatiotemporally ordered passive membranes causes entropy reduction of Ti open system. The stable chemisorbed monolayer starts metallic passivation and then relying on a passivation mechanism of incremental oxidation damping the underlying TiOx@Ti ceramet-like gradient oxide layer formed. Both layers work together to effect passivation. The entropy of Ti open system is defined as *S0*. Once contacted with environmental oxygen, the entropy of Ti is defined as *S1*. The entropy of initial passivation state of Ti is defined as *Si*, while that of further passivation

state as $Sf$. $\Delta S=Sf-Si=d_iS1+d_iS2+d_iS3+d_eS1+d_eS2<0$. (**C**) Illustration of passivation mechanism of incremental oxidation damping. Through the stable chemisorbed monolayer, the adsorbed environmental oxygens such as $O_2$, $O_3$ and O atoms are converted into $O^{2-}$ ions. And the underlying oxide layer is able to absorbs the $O^{2-}$ ions in diffusion to continuously fortify gradient-ordered structure of Ti passive membrane via breaking symmetry in the direction of oxidation. The growth of Ti passive membrane effects passivation. $t_0$: formation time of the chemisorbed Ti-O monolayer; $t_1$: initial formation time of Ti passive membrane.

# Supplementary Materials for

## The Growth of Passive Membranes: Evidence from Ti and Like Metals

Weiwei Lao, Qiaojie Luo, Ying Huang, Haixu Zhong, Chaoqian Lou, Xuliang Deng, Xiufang Wen, Xiaodong Li

Corresponding author: cisarli@zju.edu.cn; kqdengxuliang@bjmu.edu.cn

**The PDF file includes:**

Materials and Methods
Figs. S1 to S12
Tables S1 to S2

## Materials and Methods

Samples treatment with three oxidizing environments

Ti discs (10 mm in diameter, 1 mm in thickness) and Ti plates (size of 10 mm × 20 mm × 1mm) were prepared by machining commercial pure Ti materials (Grade I). Ground and mechanochemical polished to 100 Å roughness. Polished samples were then successively ultrasonically cleaned with acetone, ethanol and deionized water, and dried with nitrogen. All discs were treated under three kinds of gradient-upgrading oxidizing environments, (1) NA-Ti group, treated under the atmospheric environment; (2) $O_3$-Ti group, treated under the atmospheric environment which contains pumped $O_3$ (42.5±5.3mg/h); (3) O-Ti group, treated under UV irradiated atmospheric environment that includes UV-split O atoms and $O_3$ (44.4±3.9mg/h) formed via the combination of O atoms and $O_2$ (formulas 1 and 2). The wavelengths of UV irradiation were 253.7nm (100%) and 184.9nm (19%) respectively, which were performed with two 8W bactericidal lamps (Cnlight, Guangdong, China) for different times in an ultraviolet disinfection cabinet with volume of 40cm*30cm*25cm.

Al, Ni, Nb, Zr and Ta plates (size of 10 mm × 10 mm × 1mm) were prepared separately by machining commercial pure materials and were treated under different oxidizing environments. (1) NA-Al/Ni/Nb/Zr/Ta group, treated under the atmospheric environment; (2) O-Al/Ni/Nb/Zr/Ta group, treated under UV irradiated atmospheric environment that includes UV-split O atoms and $O_3$ as described in the previous text.

According to the Chapman Mechanism, when the UV wavelength is below 242nm, following reactions occur in air:

$$O_2 \xrightarrow{h\gamma} [O] + [O] \qquad (1)$$

$$[O] + O_2 \longrightarrow O_3 \qquad (2)$$

Herein, $h\gamma$ is the UV photon and [O] is a free oxygen atom.

Chemical composition measurements by X-ray photoelectron spectroscopy

The chemical composition of the surfaces was evaluated by X-ray photoelectron spectroscopy (AXIS Supra, Kratos, UK) with *Al Kα* monochromatic X-ray source ($h\nu = 1486.6\ eV$) under high vacuum conditions. Relative concentrations of the detected elements were calculated by taking the percentages of their peak-to-peak heights in differentiated survey spectra, and after correction by atomic sensitivity factors.

Ultra-violet photoelectron spectroscopy study

Ultra-violet photoelectron spectroscopy (UPS) with a pass energy of 1 eV, a target current of 50 mA was used to calculate the work function of the materials. Here, He I (21eV) was used to turn on the UV source. Work function (φ) is a parameter signifying the electronic stability of the crystallographic structure and its theoretical formula is:

$$\varphi = Ee - E_F \qquad (3)$$

Herein, Ee is the potential energy of electrons outside the surface of Ti passive film and $E_F$ the Fermi energy level of the solid material. Generally, the potential of electrons on the vacuum side outside the material surface is selected as the reference zero point, that is, the vacuum energy level is set as Evac = 0, and Ee = Evac =0. Therefore, the above formula is:

$$\varphi = -E_F \qquad (4)$$

Computational methods and models

The adsorption models of O atoms on different crystal faces of hcp Ti as well as oxide models of TiO were constructed respectively.

The Density functional theory (DFT) calculations of adsorption in models were conducted using the Cambridge Sequential Total Energy Package (CASTEP) computer code. The Perdew-Burke-Ernzerhof functional within the generalized gradient approximation (GGA-PBE) was generally regarded as the most appropriate for the investigation of molecular adsorbate systems and applied to acquire the electronic energies (*50*). The cutoff energy of kinetic energy was set to 571.4eV and the width of smearing was set to 0.1eV with 4×4×1 mesh of k-points. Meanwhile, 5-layer slabs of Ti atoms were built, and all crystal surfaces of hcp Ti were used to establish the p (2×2) supercell. There was a 15 Å vacuum slab to avoid the interaction between periodically repeated slabs (*51, 52*).

TEM specimen preparation and technology

The cross-sectional TEM specimen (10μm in width) was prepared by an FEI Quanta 3D FEG in combination with focused ion beam (FIB)/SEM workstation with 5kV working voltage of the electron beam and 2-30kV voltage of ion beam. As a result, we obtained a distinct metal/oxide film interface and characterize the structure of the interface region. During sample preparation and subsequent TEM observation, the thin passive film was strictly ensured free of mechanical and beam-induced damage.

TEM observation

The HRTEM and HAADF-STEM images were obtained by aberration-corrected transmission electron microscopy (Titan $G^2$ 80-200 ChemiSTEM, FEI, USA) at a working voltage of 200kV in order to observe micro morphologies and phase structures. Selected area electron diffraction (SAED) was used to certify crystal structures of the passive film and its adjacent areas. We also used the advanced Super-X EDS system with four detectors for the mapping and line scan analysis to evaluate element contents in different depths of the passive film, which significantly shortened the experimental span and thus effectively avoided beam damages to specimens. The Electron energy loss spectroscopy (EELS) analysis was performed in a FEI Titan $G^2$ 80-200 ChemiSTEM operated at 200kV with a nominal energy resolution of 1.1eV. Under the STEM mode, the convergence angle was 21.4mrad and the collection angle was 38.07mrad. EELS data were only smoothed after deducting the background.

Chemical phase detection by thin film X-ray diffraction

The phase composition was studied by thin film X-ray diffraction (XRD) using an X-ray diffractometer (D8 DISCOVER, Bruker, Germany) with Cu Ka1 (λ= 1.540598 Å)/Ka2 (λ=1.544426 Å) radiation with a ratio of 0.5, operating at 40 kV and 25 mA with a scan step size of 0.0167113° between 5° and 80° of measured scan.

After measuring the diffraction pattern, the Williamson Hall (WH) method was used for further analysis. In WH plots, the full width at half maximum (FWHM) is plotted relative to the diffraction angle for each diffraction peak. This method is the basic approach to evaluate the micro-strain $\varepsilon$ that is produced by dislocations[49]. When X-rays with a wave length $\lambda$ are used for diffraction analysis and a diffraction angle $\theta$ and FWHM $\beta$ is obtained in each diffraction peak, the following Williamson-Hall (WH) equation is constructed as follows:

$$\frac{\beta\cos\theta}{\lambda} = \alpha + \frac{\varepsilon \cdot 2\sin\theta}{\lambda} \quad (5)$$

Here, the parameter α is dependent on the crystallite size. In poly crystalline titanium metals, several diffraction peaks appear and $\frac{\beta\cos\theta}{\lambda}$ and $\frac{2\sin\theta}{\lambda}$ are obtained from each diffraction peak. Therefore, the values of parameter α and ε are determined based on the relation between $\frac{\beta\cos\theta}{\lambda}$ and $\frac{2\sin\theta}{\lambda}$. We further calculated $\frac{\beta\cos\theta}{\lambda}$ and $\frac{2\sin\theta}{\lambda}$ from every diffraction peak of tested XRD curves and made a linear fit of them. The slope of each curve represents the average micro-strain ε of each Ti surface.

Electrochemical measurements

Ti samples in size of 10 mm × 20 mm × 1mm were used for electrochemical measurements. A classical three-electrode cell with titanium plate as study electrode, a Pt wire as counter electrode and Ag/AgCl as reference electrode was connected to a CS310H electrochemical workstation (Wuhan Corrtest Instruments Corp., Ltd., CHN), and a computer with CS Studio5 software was used for all the electrochemical measurements. A PBS (NaCl: 145mM, $Na_2HPO_4 \cdot 12H_2O$: 8.1mM, $NaH_2PO_4 \cdot 2H_2O$: 1.9mM) solution with pH 7.2-7.4 was used as electrolyte. The analyzed area immersed in electrolyte was 10 mm×10 mm of the titanium surface. All potential data in the following are refer to Ag/AgCl electrode and all electrochemical measurements were carried out at 37±1°C temperature. The open circuit potential (OCP) measurement was carried out for 30 min starting from the electrode immersing into the electrolyte. The electrochemical impedance spectroscopy (EIS) measurements were carried out under potentiostatic condition at OCP with 10mV amplitude AC voltage signal, and the applied frequency range was from $10^5$Hz down to $10^{-2}$ Hz. The impedance date was analyzed using ZSimpWin 3.0 software. The potentiodynamic polarization curves were obtained in the range of OCP ± 0.5V using a scan rate of 1 mV/s. The corrosion potential ($E_{corr}$) and corrosion current density ($I_{corr}$) were determined by Tafel slope extrapolation. All electrochemical tests were repeated three times to ensure reproducibility and statistically analyzed to gain the standard deviations.

Electrochemical passivation treatments

The electrochemical polarization to the titanium specimens (10 mm × 20 mm × 1mm) was carried out in 1 M NaOH solution and 1 M $H_2SO_4$ solution at room temperature (25±1°), respectively. A classical three-electrode cell with titanium plate as the working electrode, a Pt wire as counter electrode and Ag/AgCl (3 mol $L^{-1}$ KCl solution) as reference electrode was connected to a CS310H electrochemical workstation (Wuhan Corrtest Instruments Corp., Ltd., CHN). All potential values are reported against the Ag/AgCl electrode in our work. Prior to the measurement, the specimen was kept at −1.0 V for 10 min in order to electrochemically reduce the oxide film formed by exposure of the titanium surface to air. The polarization was then implemented by scanning in the reverse direction to a proper potential so that the entire passive region was presented. The scan rate was set to be 5 mV/s. In 1 M $H_2SO_4$ solution, the potentiostatic polarization of the titanium sample for further analysis was carried out at 1.0 V (P1) for 9h when the current density was stable. While in 1 M NaOH, potentiostatic polarization measurement was carried out at 1.5 V (P2) for 9h (as shown in fig. S12). On completion of the electrochemical passivation, the titanium disk was taken out of the cell rapidly followed by thorough rinsing with deionized water, and then dried 30 min in a vacuumed desiccator at room temperature.

Statistical analysis

Relative contents of elements from XPS data in accordance with normal distribution and homogeneity (Shapiro–Wilk Test, $p > 0.05$ & Levene Test, $p > 0.05$) were compared by two-way ANOVA test followed by Tukey's Post hoc test. All data are shown as mean ± standard deviation (SD) and have been tested using SPSS statistical software (V20, SPSS, Inc., USA). The statistical significance (p value) was set at a level of $p=0.05$.

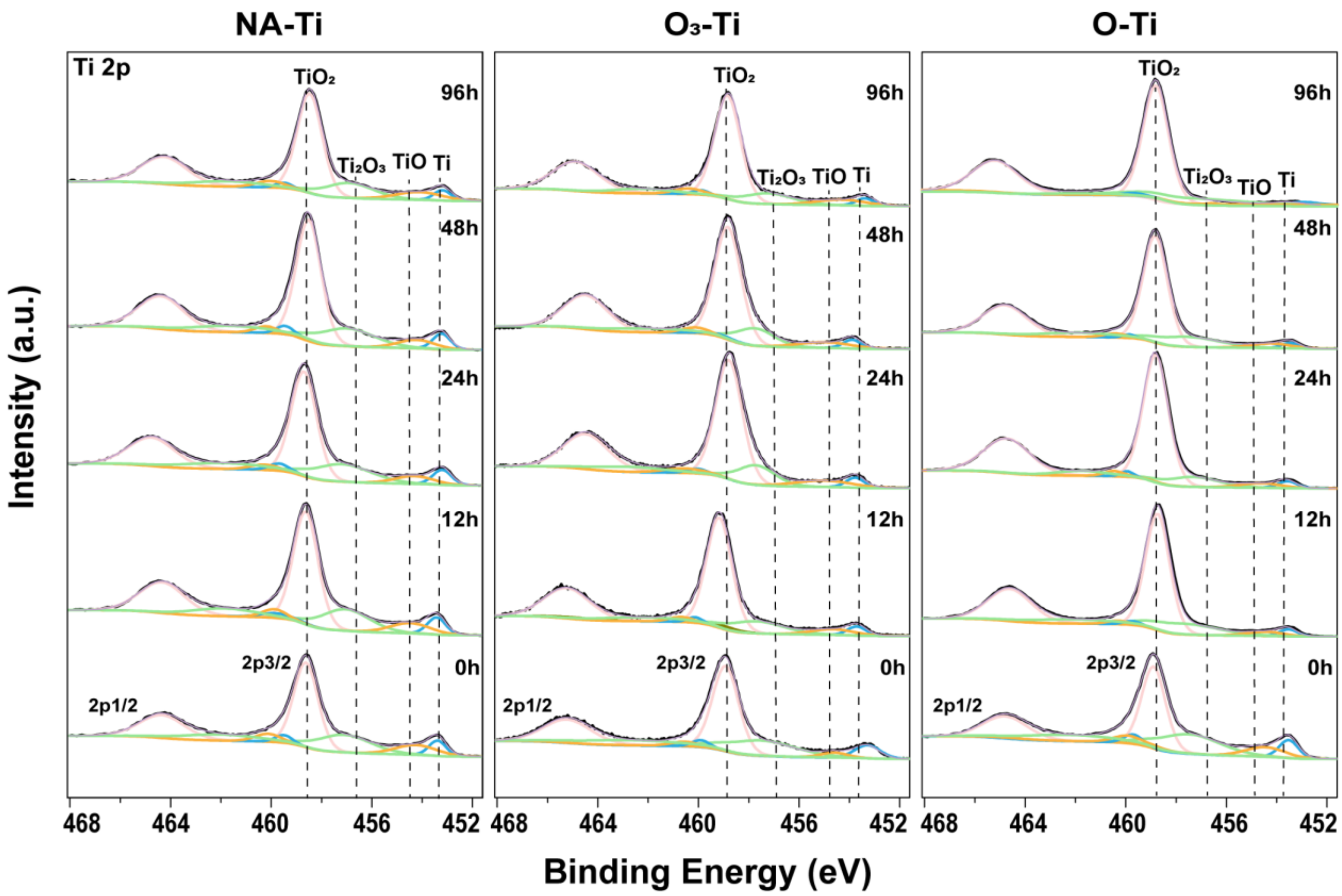

**Fig. S1. Elemental analysis of XPS spectra on three Ti passive membranes.** Deconvolution of $Ti^{2p}$ peaks in XPS spectra on Ti passive membranes of NA-Ti, $O_3$-Ti and O-Ti with treatment of 0~96h.

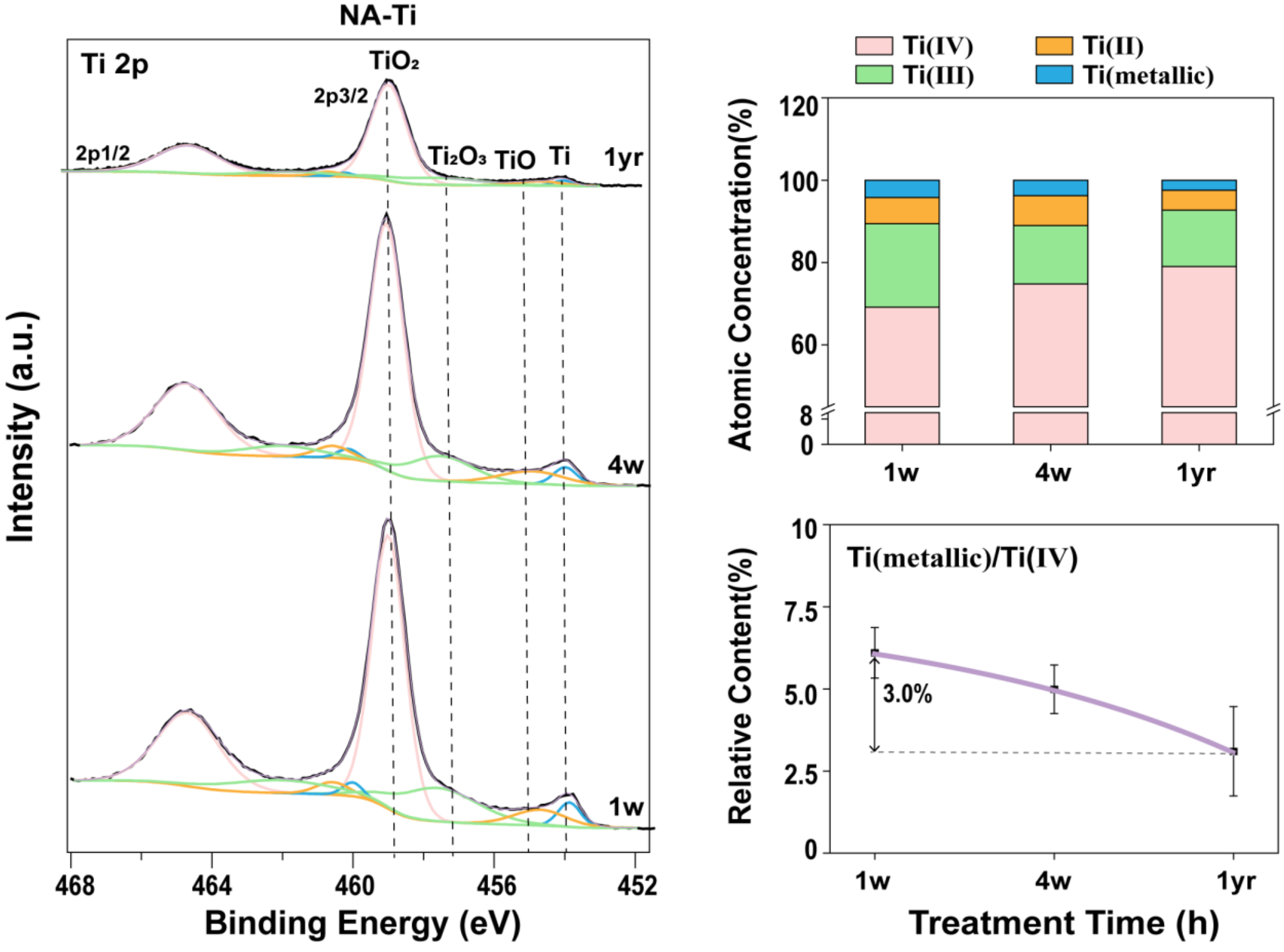

**Fig. S2. Elemental analysis of XPS spectra on Ti passive membranes of NA-Ti for aging with different times.** Left: Deconvolution of $Ti^{2p}$ peaks in XPS spectra on Ti passive membranes for aging one week, four weeks and one year. Right upper: content percentage of Ti(IV), Ti(III), Ti(II) and Ti(metallic); Right down: trend charts of $Ti^{metallic}/Ti^{4+}$ relative content. *$p < 0.05$ vs. treatment time of one week in the same group.

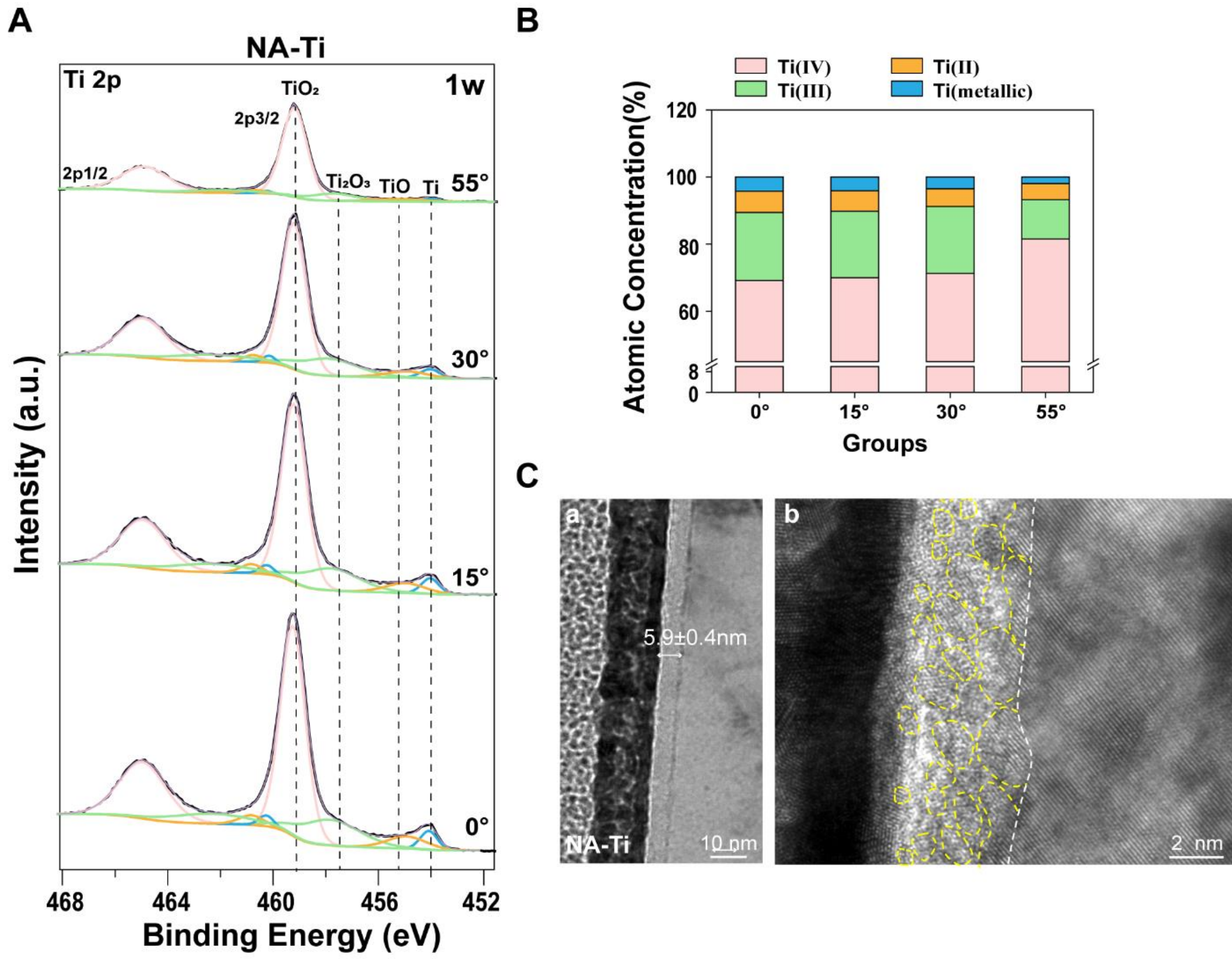

**Fig. S3. XPS spectra and HRTEM observation on the Ti passive membrane of NA-Ti for aging with one week.** (**A**) Angle-resolved XPS spectra of Ti2p. (**B**) Content percentage of Ti(IV), Ti(III), Ti(II) and Ti(metallic) at different take off angles. (**C**) Thickness and high-resolution structures of the passive membrane [(a) and (b)].

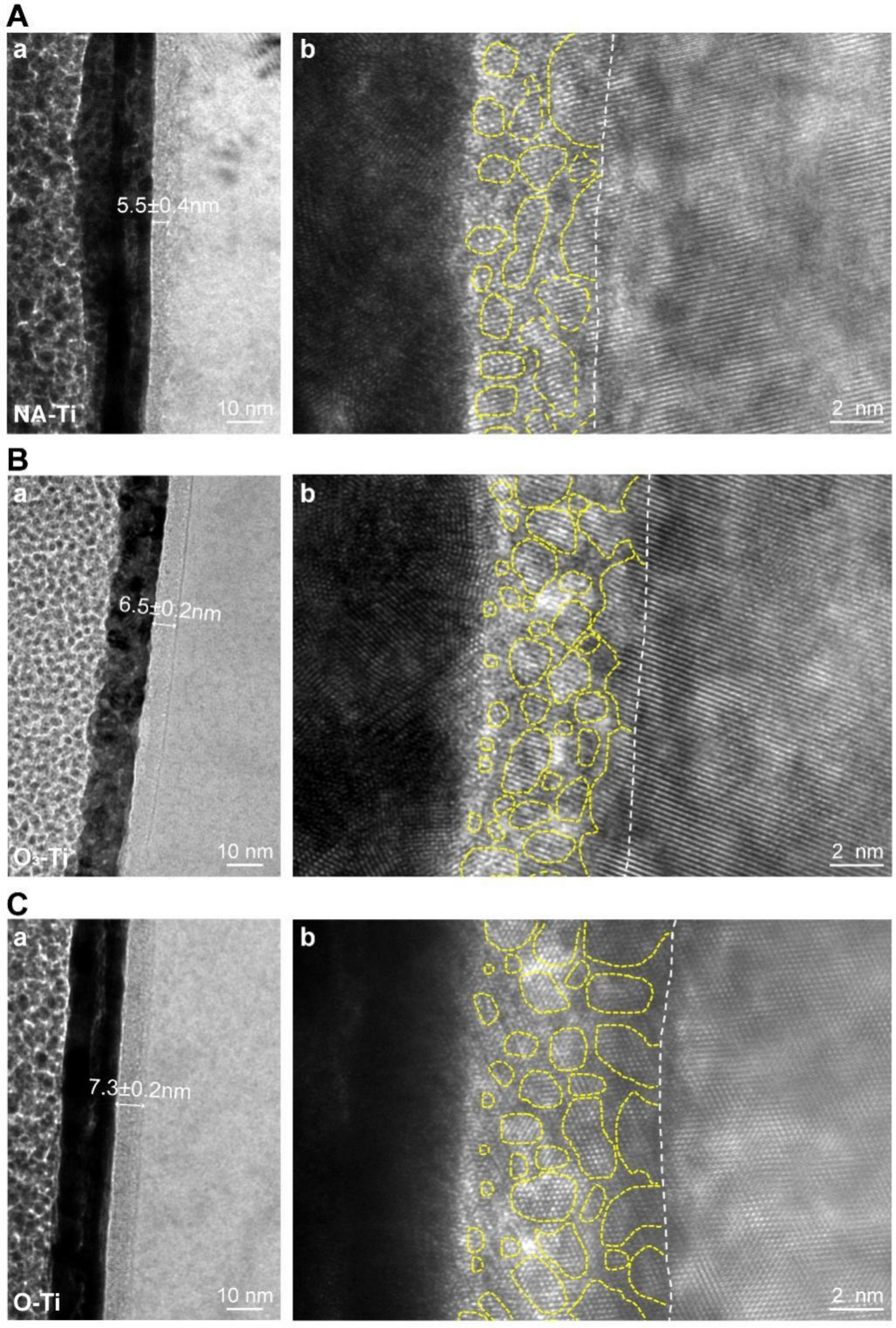

**Fig. S4. HRTEM observations of three Ti passive membranes.** Passive membranes on O-Ti and $O_3$-Ti are thicker than that on NA-Ti with treatment for 48h [**A**(a), **B**(a) and **C**(a)]. Under room temperature atmosphere, Ti passive membrane shows a TiOx@Ti cermet-like gradient structure of high entropy [**A**(b)]. From the upper to the bottom, with the amorphous oxides gradually withering, the Ti lattice gradually increase until reverting to Ti matrix. Under stronger oxidizing environment ($O_3$ or O), thicker passive membranes with more amorphous oxides are induced to protect the internal matrix from stronger oxidation [**B**(b) and **C**(b)].

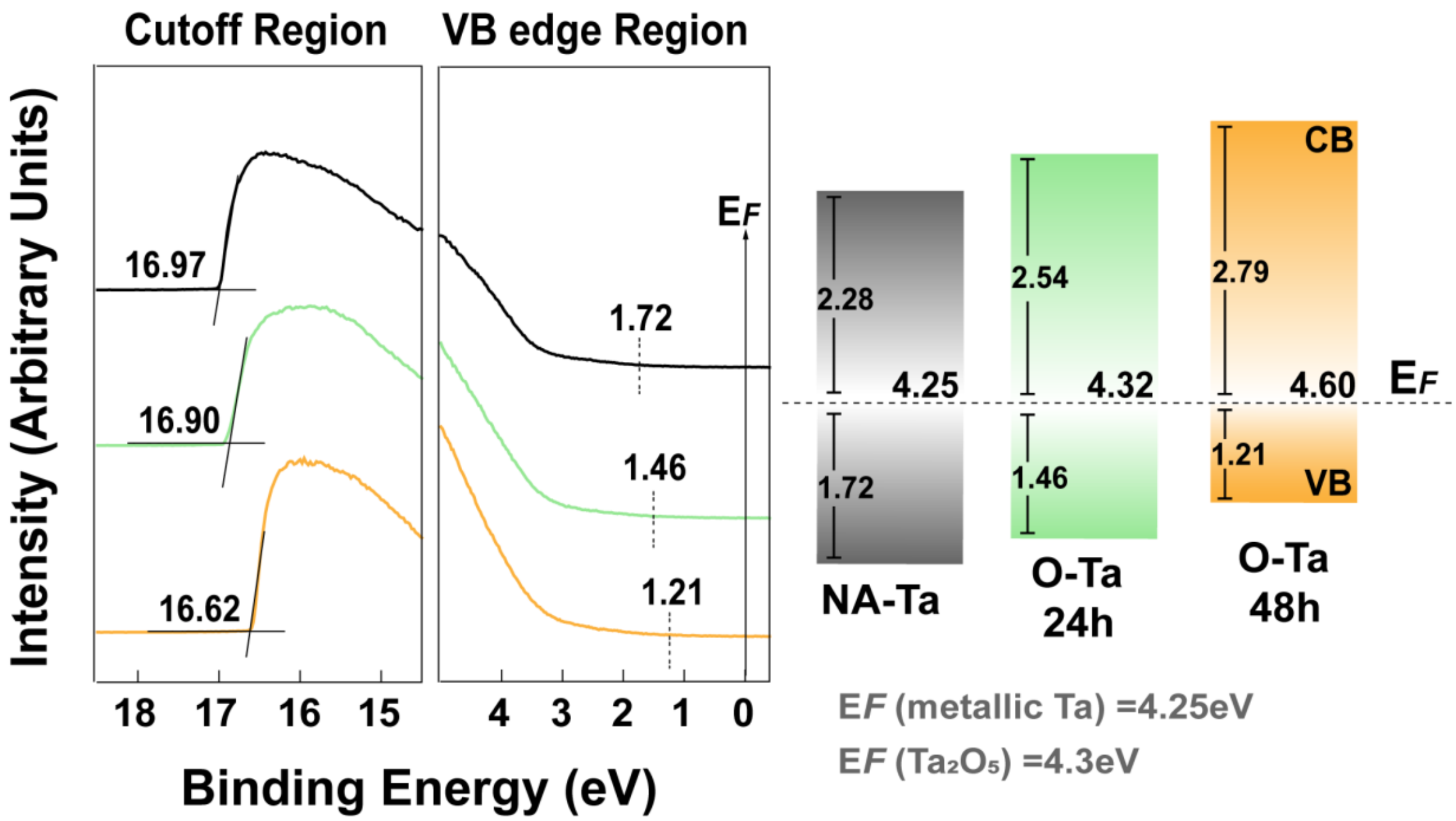

**Fig. S5. UPS spectra of the cutoff region and VB edge region for Ta passive membranes.** Spectra from bottom to top represent passive membranes of NA-Ta, O-Ta with treatment for 24h and 48h. Right three columns represent correlative UPS measured energetic levels. $E_F$, CB, and VB represent Fermi level, conductive band and valence band, respectively.

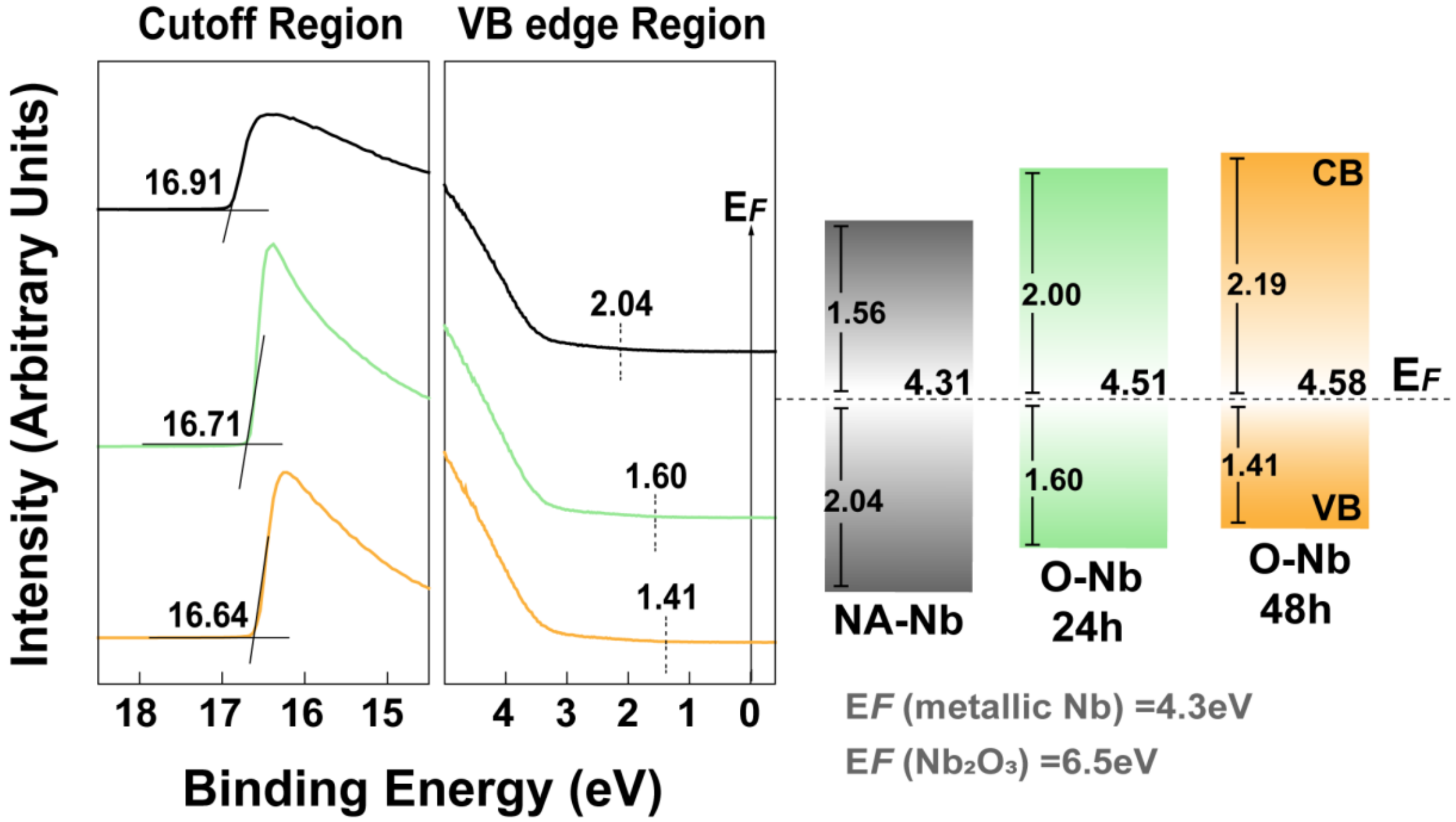

**Fig. S6. UPS spectra of the cutoff region and VB edge region for Nb passive membranes.** Spectra from bottom to top represent passive membranes of NA-Nb, O-Nb with treatment for 24h and 48h. Right three columns represent correlative UPS measured energetic levels. $E_F$, CB, and VB represent Fermi level, conductive band and valence band, respectively.

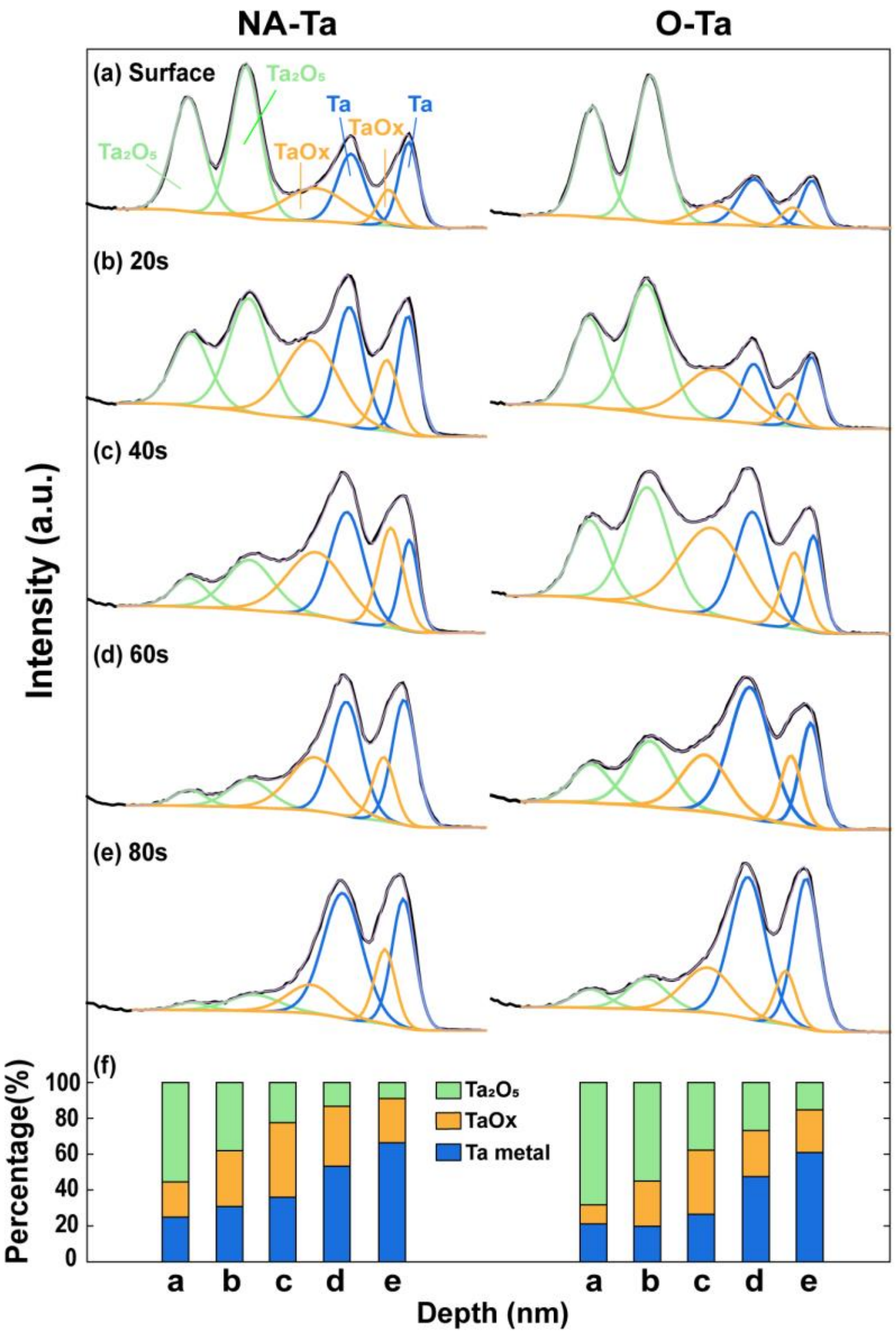

**Fig. S7. Ta passive membranes in a gradient-order state.** Deconvoluted Ta4f XPS spectra of (a) 0s sputter time, (b) 20s sputter time, (c) 40s sputter time, (d) 60s sputter time and (e) 80s sputter time of the passive membrane formed on Ta metal and (f) percentages of the component peaks to the total intensity of Ta4f at different depths of NA-Ta and O-Ta, respectively.

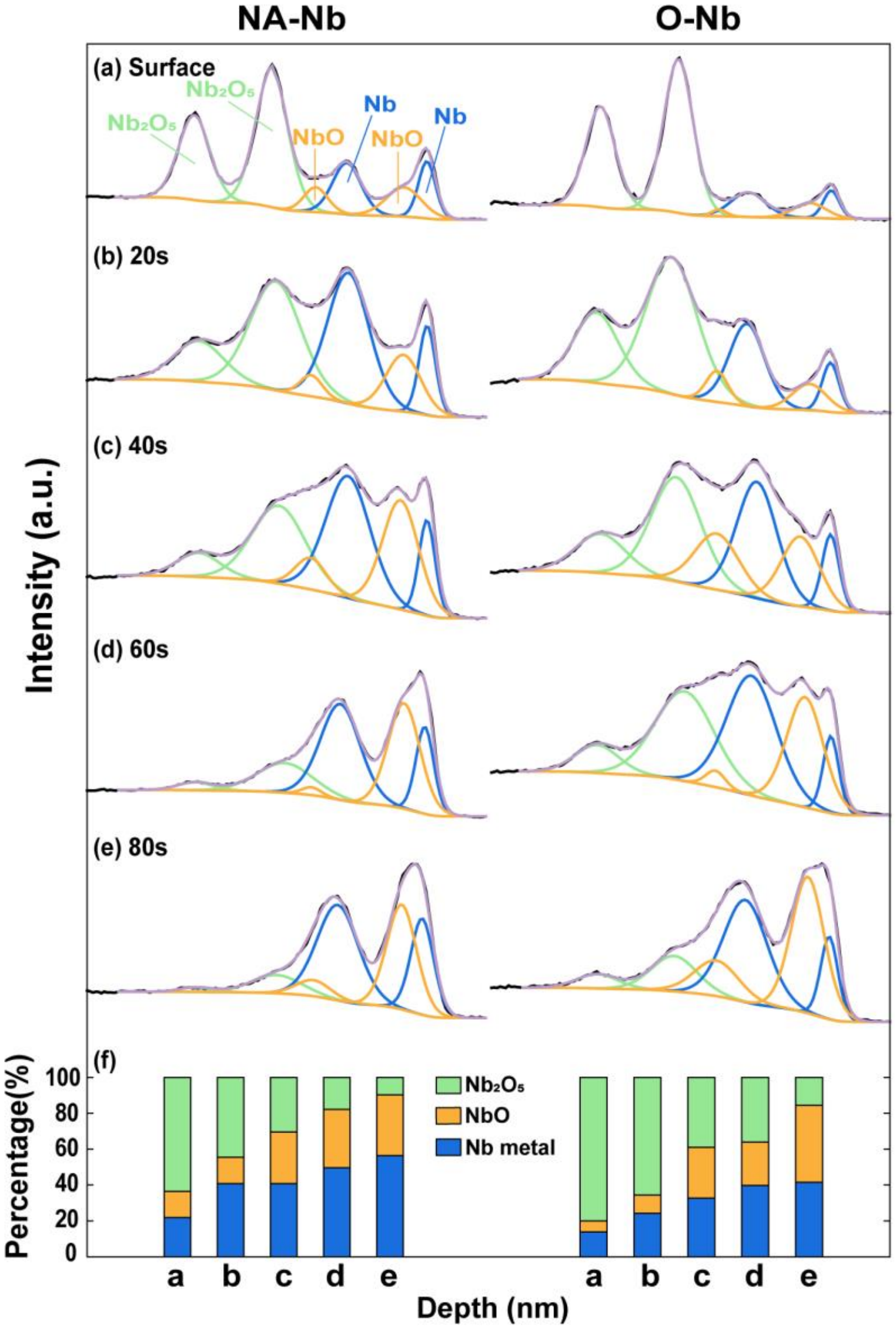

**Fig. S8. Nb passive membranes in a gradient-order state.** Deconvoluted Nb3d XPS spectra of (a) 0s sputter time, (b) 20s sputter time, (c) 40s sputter time, (d) 60s sputter time and (e) 80s sputter time of the passive membrane formed on Nb metal and (f) percentages of the component peaks to the total intensity of Nb3d at different depths of NA-Nb and O-Nb, respectively.

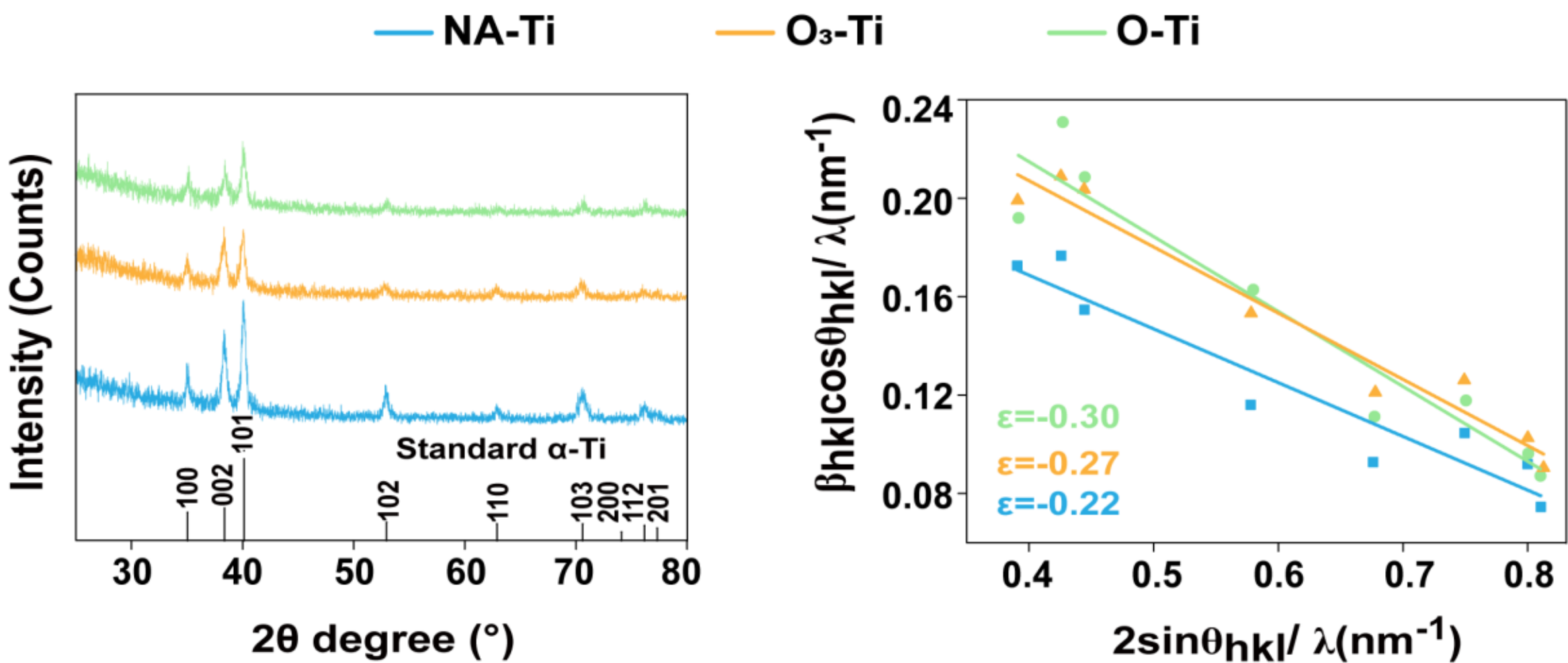

**Fig. S9. Micro-strains and O-vacancies analysis of three Ti passive membranes.** The micro-strain $\varepsilon$ of each corresponding Williamson-Hall (WH) plot (right) from TF-XRD results (left) showed that there are more structural defects in oxide layers of O-Ti and $O_3$-Ti than NA-Ti with treatment for 48h.

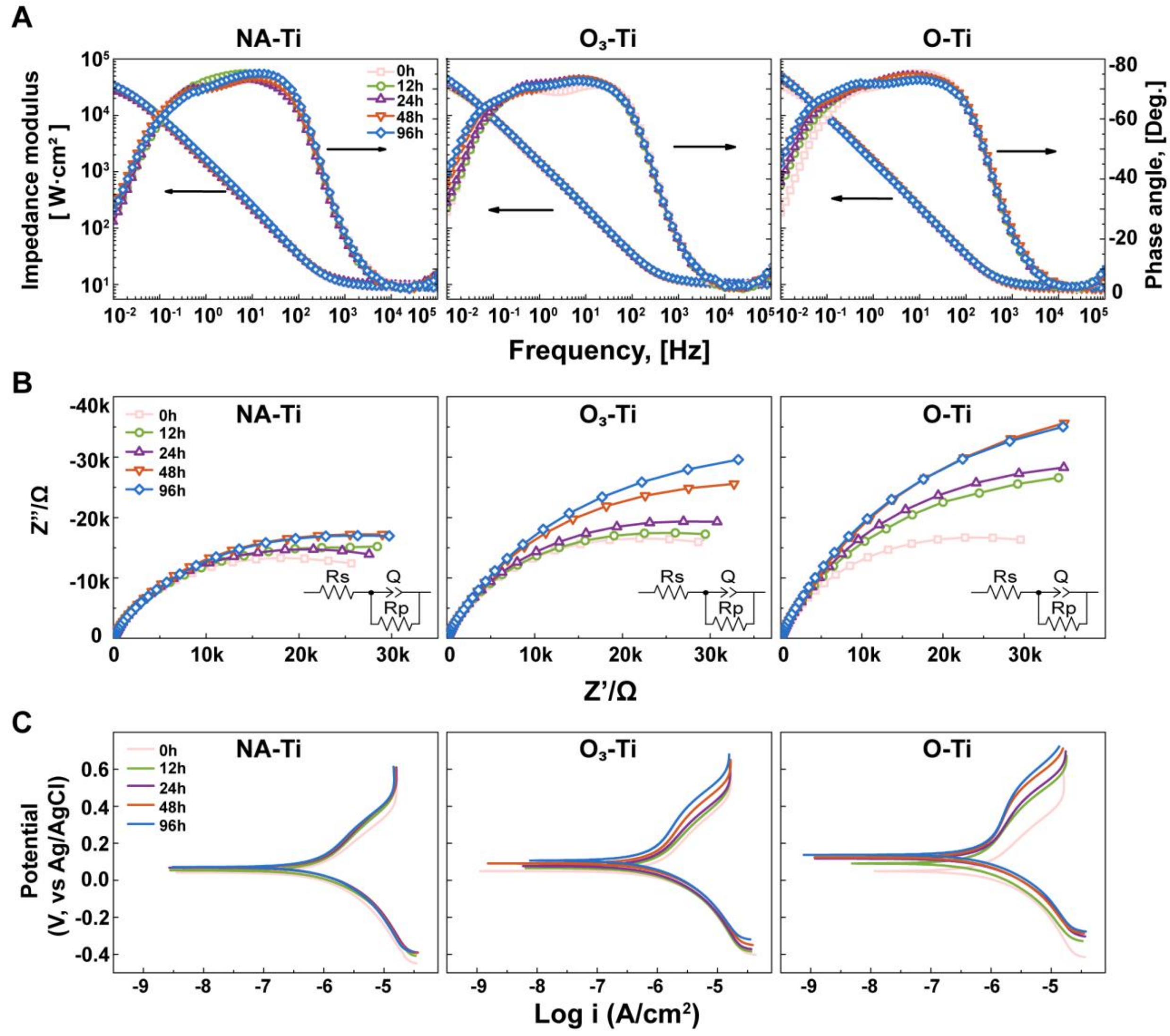

**Fig. S10. Electrochemical measurements of the three Ti passive membranes.** (**A**) Bode plots from EIS of NA-Ti, $O_3$-Ti and O-Ti. (**B**) Nyquist plots from EIS of NA-Ti, $O_3$-Ti and O-Ti. Inset is the equivalent circuit of membrane/solution interface. The impedance behavior of the passive membrane can be described by the equivalent circuit of the membrane/solution interface, *Rs* (*QRp*). In the equivalent circuit, *Rs* represents the resistance of solution; *Q* is the constant phase elements (CPE) used for describing the capacitance behaviors of the compact passive membrane and *Rp* is the polarization resistance. (**C**) Potentiodynamic polarization curves of NA-Ti, $O_3$-Ti and O-Ti.

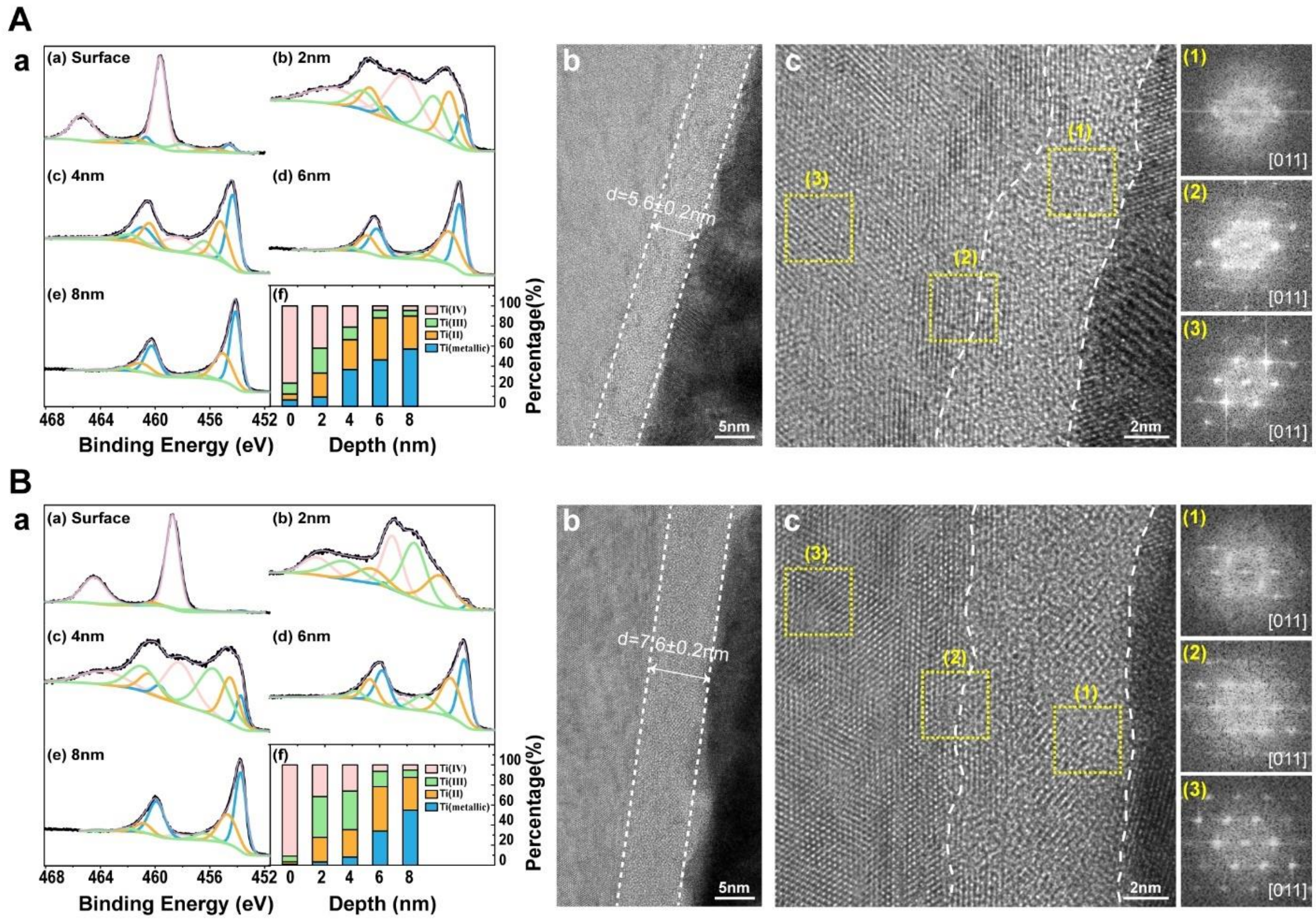

**Fig. S11. Structures of the electrochemically-made Ti passive membranes.** (**A**) An HR-TEM image of the titanium surface after passivation in 1 M $H_2SO_4$ solution at 1.0V for 9h [**A**(a)]. The FFT patterns taken from selected regions in the passive membrane (1), transition layer (2) and Ti matrix layer (3) [**A**(b)]. Deconvoluted Ti2p XPS spectra of (a) surface, (b) 2nm, (c) 4nm, (d) 6nm and (e) 8nm of the passive membrane formed on titanium and (f) percentages of the component peaks to the total intensity of Ti2p at different depths [**A**(c)]. (**B**) An HR-TEM image of the titanium surface after passivation in 1 M NaOH solution at 1.5V for 9h [**B**(a)]. The FFT patterns taken from selected regions in the passive membrane (1), transition layer (2) and Ti matrix layer (3) [**B**(b)]. Deconvoluted Ti2p XPS spectra of (a) surface, (b) 2nm, (c) 4nm, (d) 6nm and (e) 8nm of the passive membrane formed on titanium and (f) percentages of the component peaks to the total intensity of Ti2p at different depths [**B**(c)].

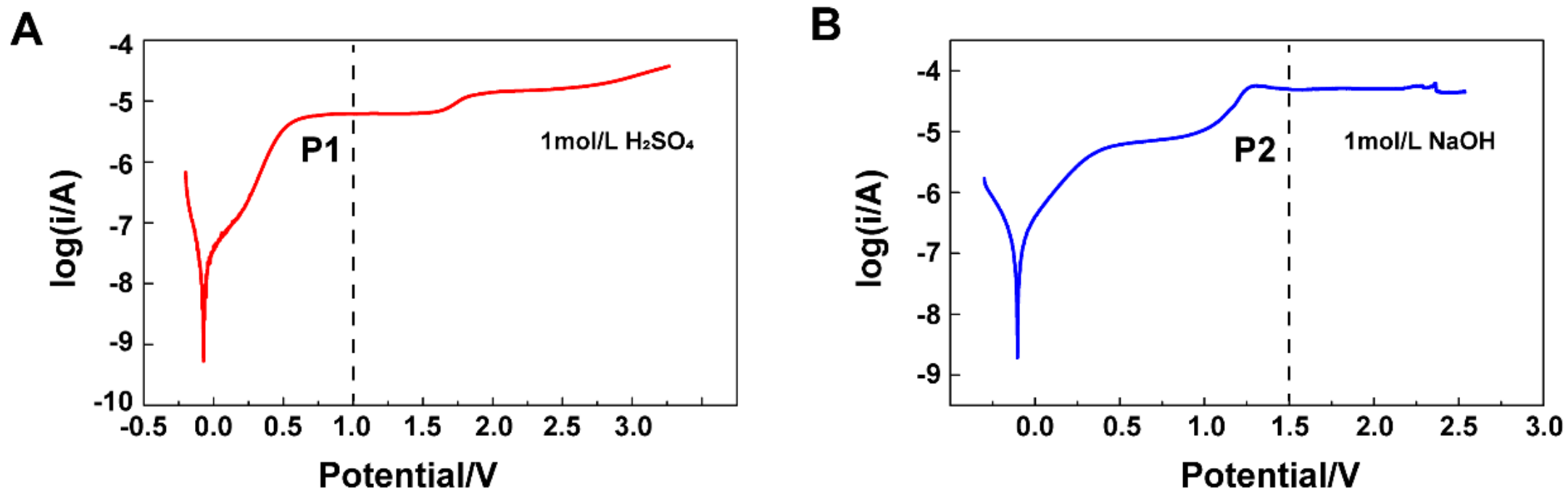

**Fig. S12. Electrochemical measurements.** Potentiodynamic polarization curves of pure titanium disks electrode in 1 M $H_2SO_4$ solution (**A**) and 1 M NaOH solution (**B**).

**Table S1. EIS parameters of different treated Ti samples**

| Groups | Treatment Time (h) | $R$s ($\Omega$ cm$^2$) | $Q$ ($\mu$F cm$^{-2}$) | n | $R_p$ (k$\Omega$ cm$^2$) | Chi squared |
|---|---|---|---|---|---|---|
| NA-Ti | 0 | 9.82±0.14 | 148.70±0.90 | 0.85±0.01 | 35.00±0.67 | $2.22\times10^{-3}$ |
| | 12 | 9.42±0.09 | 132.80±1.95 | 0.86±0.01 | 37.02±1.47 | $2.79\times10^{-3}$ |
| | 24 | 8.60±0.18 | 147.13±2.48 | 0.85±0.02 | 37.46±1.43 | $2.21\times10^{-3}$ |
| | 48 | 8.84±0.12 | 132.97±0.95 | 0.84±0.01 | 40.97±0.29 | $5.27\times10^{-3}$ |
| | 96 | 9.17±0.44 | 143.77±1.80 | 0.84±0.01 | 43.51±0.84 | $3.43\times10^{-3}$ |
| $O_3$-Ti | 0 | 9.60±0.32 | 154.60±3.61 | 0.83±0.02 | 44.23±2.06 | $3.68\times10^{-3}$ |
| | 12 | 10.41±0.50 | 147.50±2.88 | 0.84±0.03 | 44.50±2.90 | $1.95\times10^{-3}$ |
| | 24 | 10.31±0.37 | 146.57±1.72 | 0.85±0.02 | 47.62±2.96 | $2.37\times10^{-3}$ |
| | 48 | 10.26±0.80 | 151.27±2.04 | 0.84±0.01 | 64.33±0.57 | $3.58\times10^{-3}$ |
| | 96 | 9.93±0.74 | 150.33±2.00 | 0.82±0.01 | 74.56±1.88 | $3.28\times10^{-3}$ |
| O-Ti | 0 | 8.27±0.23 | 140.73±1.21 | 0.84±0.02 | 41.93±1.68 | $3.58\times10^{-3}$ |
| | 12 | 9.27±0.20 | 136.27±3.19 | 0.84±0.02 | 62.84±1.50 | $3.46\times10^{-3}$ |
| | 24 | 8.86±0.13 | 139.70±2.20 | 0.83±0.03 | 67.49±1.56 | $3.04\times10^{-3}$ |
| | 48 | 8.28±0.27 | 140.10±2.19 | 0.84±0.03 | 88.82±3.18 | $3.58\times10^{-3}$ |
| | 96 | 8.84±0.21 | 146.97±2.34 | 0.83±0.02 | 92.53±1.74 | $3.08\times10^{-3}$ |

**Table S2. Corrosion parameters of potentiodynamic polarization for different treated Ti samples in PBS solution at 37°C**

| Groups | Treatment Time (h) | $E_{corr}$(mV) | $I_{corr}$ (μA $cm^{-2}$) |
|---|---|---|---|
| NA-Ti | 0 | 101.41±1.53 | 1.26±0.06 |
| | 12 | 101.52±0.81 | 1.23±0.15 |
| | 24 | 103.72±1.33 | 1.22±0.20 |
| | 48 | 106.94±1.79 | 1.20±0.17 |
| | 96 | 107.83±0.76 | 1.18±0.04 |
| $O_3$-Ti | 0 | 112.12±0.47 | 1.28±0.04 |
| | 12 | 114.53±3.15 | 1.02±0.07 |
| | 24 | 119.65±2.43 | 0.89±0.03 |
| | 48 | 123.63±1.26 | 0.76±0.02 |
| | 96 | 129.63±0.55 | 0.70±0.05 |
| O-Ti | 0 | 127.65±1.96 | 1.51±0.09 |
| | 12 | 133.67±2.53 | 0.66±0.04 |
| | 24 | 139.53±1.24 | 0.64±0.05 |
| | 48 | 142.38±2.62 | 0.63±0.03 |
| | 96 | 149.62±1.21 | 0.60±0.01 |